%% file: main.tex
\documentclass[%
reprint,
superscriptaddress,
amsmath,amssymb,
aps,
prx
]{revtex4-2}

\usepackage{graphicx}
\usepackage{dcolumn}
\usepackage{bm}
\usepackage[colorlinks]{hyperref}
\usepackage{physics}
\usepackage[squaren]{SIunits}
\usepackage{amsmath,amssymb,amsthm,mathtools}
\usepackage{xcolor}
\usepackage{mathrsfs}
\usepackage{times}
\usepackage[normalem]{ulem}
\usepackage{algorithm2e}
\usepackage{xurl}
\usepackage{comment}

\usepackage{appendix}
\usepackage{titletoc}
\usepackage{tocloft}
\usepackage{etoolbox}
\usepackage{titlesec}
\usepackage{textcase}

\usepackage{tikz}
\usepackage{quantikz}
\tikzset{
  operator/.append style={
    minimum height=0.85cm,
    inner ysep=0pt
  }
}

\usepackage{array}
\usepackage{booktabs}
\usepackage{enumitem}
\usepackage{tikz}
\usetikzlibrary{arrows.meta,calc,fit,positioning}
\usepackage[nameinlink,capitalize]{cleveref}

\newcommand{\F}{\mathbb{F}}
\newcommand{\bits}{\{0,1\}}
\newcommand{\diag}{\operatorname{diag}}
\newcommand{\im}{\operatorname{im}}
\newcommand{\supp}{\operatorname{supp}}
\newcommand{\Unif}{\operatorname{Unif}}
\newcommand{\TV}{d_{\mathrm{TV}}}
\newcommand{\Alt}{\operatorname{Alt}}

\DeclareMathOperator{\Law}{Law}
\newcommand{\Win}{\mathsf{Win}}
\newcommand{\Ball}[2]{B_{#1}(#2)}

\definecolor{deepblue}{HTML}{173B57}
\definecolor{softblue}{HTML}{EAF2F8}
\definecolor{softgray}{HTML}{F3F5F7}
\definecolor{darkgray}{HTML}{4A4F55}
\definecolor{accent}{HTML}{9B3A31}

\newtheorem{theorem}{Theorem}
\newtheorem{lemma}[theorem]{Lemma}
\newtheorem{corollary}[theorem]{Corollary}
\newtheorem{proposition}[theorem]{Proposition}

\newtheorem{definition}[theorem]{Definition}

\hypersetup{
	colorlinks=true,
	linkcolor=blue,
	filecolor=blue,      
	urlcolor=blue,
	citecolor=blue
}
\newcommand{\xg}[1]{{\color{magenta}{#1}}} 

\begin{document}

\title[]{Quantum-Classical Separation in Probability Distribution Transport via Nonlocal Games}

\makeatletter

\author{Fangjun Hu}
\email{fhu@quera.com}
\affiliation{QuEra Computing Inc., 1380 Soldiers Field Road, Boston, MA 02135, USA}

\author{Xun Gao}
\email{xun.gao@colorado.edu}
\affiliation{JILA and Department of Physics, University of Colorado Boulder, Boulder, CO 80309, USA}

\date{\today}

\begin{abstract}
    In many-body physics and generative artificial intelligence, locality provides a natural notion of phase equivalence: two quantum states or classical probability distributions are regarded as in the same phase when local dynamics can reversibly transform one into the other.
    This raises a fundamental question: can local quantum dynamics transport classical probability distributions in ways that local classical dynamics cannot?
    We establish a quantum-classical separation in distribution transport: two classical distributions can be connected exactly and reversibly by a constant-depth local quantum circuit, whereas classical stochastic transport requires a substantially larger spatial range.
    We demonstrate this separation explicitly on a one-dimensional chain and develop a general construction principle on graphs, based on nonlocal games.
    Moreover, the source distribution can itself be generated by a simple local classical process, establishing a quantum advantage in distribution generation as well.
    Our results reveal a fundamental distinction between classical and quantum local reversibility, even when both the initial and final states are entirely classical.
\end{abstract}

\maketitle

\section{Introduction}
\label{sec:intro}

Locality plays a central role in understanding of quantum many-body physics \cite{wen_1989_vacuum, wen_1990_topological, wen_2007_quantum, chen_2010_local, haah_2011_local, haah_2016_invariant}. 
More recently, locality in the dynamics of probability distributions has also attracted growing attention in generative artificial intelligence, particularly in diffusion models, which generate data by transporting a simple noise distribution through a sequence of approximately local denoising steps \cite{hu_2025_local, hu_2026_learning, zhang2026concurrence, liu2026breakdown, kamb2025analytic, daras2025ambient, lukoianov2025locality}.
In the traditional topological phase theory of quantum pure states, two states are often regarded as belonging to the same phase when they can be transformed into one another by local unitary operations of bounded depth or finite time \cite{chen_2010_local, zeng_2019_quantum}.
Similar ideas have recently been extended to mixed quantum states \cite{coser_2019_classification, sang_2025_statbility, sang_2025_mixed} and to classical probability distributions \cite{hu_2025_local}, where \textit{local reversibility} provides a natural notion of phase equivalence. 
Intuitively, local reversibility means that the effect of each local operation along a transportation from one mixed state (or classical distribution) to another can be undone by another local operation. Two mixed states are defined as in the same phase if there is a local and shallow transport circuit between them; moreover, all channel gates during this transportation satisfy the local reversibility.

On the other hand, diffusion models generate complicated data distributions by learning to reverse a gradual evolution toward a simple, easily sampled distribution \cite{sohl_2015_deep, song_2019_generative, song_2021_sde, anderson_1982_reverse}. 
Generically, such a transport evolution can be implemented either by classical stochastic circuits or by quantum circuits acting on the corresponding diagonal density matrices \cite{hu_2025_local, liu_2025_measurement, zhang_2024_generative}.
This raises a basic question: can quantum and classical local dynamics induce genuinely different notions of local reversibility between classical probability distributions?
More concretely, can two classical distributions be reversibly transported by shallow and local quantum dynamics while even one direction of the corresponding classical transport requires expensive macroscopic range?

In this work we show that the answer is yes.
We construct two families of classical distributions, $Q$ and $P$, defined on an $n$-bit open one-dimensional chain, which can be transformed into one another exactly by a constant-depth \textit{bidirectionally geometrically local} quantum circuit.
By contrast, any classical stochastic circuit that transforms $Q$ into $P$ with sufficiently small error must have geometric range $\Omega(n)$.
To be more specific, we use the same notion of bidirectionally geometric locality for classical and quantum circuits: the backward lightcone of every output and the forward lightcone of every input are both at most radius-$d$.
We call such a circuit geometrically local when $d=O(1)$ is independent of the system size.

The mechanism behind this separation of our construction can be understood through a nonlocal game \cite{Bravyi2018, Bravyi2020NoisyShallow, Daniel2022CyclicCluster, grier_2026_quantum} embedded in an open one-dimensional chain.
A few well-separated bits, called \textit{question bits}, play the role of nonlocal game questions, and the target distribution imposes question-dependent correlations among the remaining bits.
A shallow quantum circuit based on cluster-state entangling gates realizes these correlations, while the corresponding winning conditions cannot be satisfied simultaneously by classical responses restricted to local information.
This all-versus-nothing obstruction extends to arbitrary local stochastic circuits, establishing the classical generation hardness of $P$.
Since $Q$ itself is generated by a local classical map, composing it with any local classical transport from $Q$ to $P$ would generate $P$ locally; hence the classical hardness of generating $P$ implies the hardness of transporting $Q$ to $P$.


From a phase-of-matter perspective, our result shows that classical and quantum local operations can induce different notions of equivalence even when the states themselves are fully classical.
Namely, two classical distributions can belong to the same quantum phase while lying in different classical phases.

We focus on a one-dimensional open chain for both experimental and pedagogical reasons.
One-dimensional geometries are natural in several leading quantum platforms, including trapped ion \cite{lanyon2013measurement, ringbauer2025verifiable}, superconducting qubits \cite{gong2019genuine}, ultracold atom \cite{mandel2003controlled, he2026orbit} and neutral-atom arrays \cite{endres2016atom, levine2019parallel, bluvstein2022processor, evered2023high}, while also making the locality structure and the underlying nonlocal-game obstruction particularly transparent.
The mechanism itself, however, is not specific to one dimension.
More generally, on a large class of high-dimensional graph supporting a suitable nonlocal game, the same construction principle can be used to encode the winning constraints into classical probability distributions, producing distributions that are reversibly connected by local quantum dynamics but separated under local classical dynamics.
We explicitly give such a generic construction in Sec.~\ref{sec:nonlocal-game-transport}.

\section{Definition of the transported distributions in 1D}
\label{sec:def_PQ}

\begin{figure}
    \centering
    \includegraphics[width=0.8\linewidth]{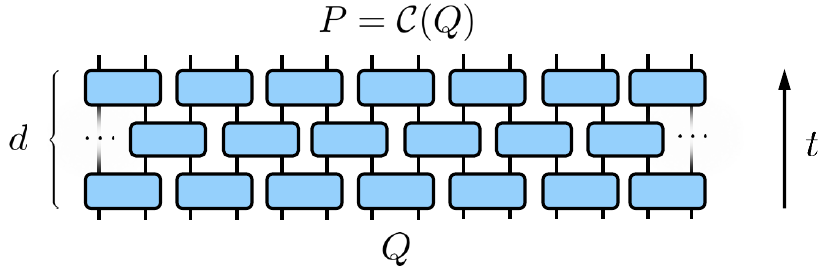}
    \caption{
    Circuit diagram of the probability transport between two distributions $P$ and $Q$ (time arrow from bottom to top).
    If all local channel gates in the circuit $\mathcal{C}$ are classical, then $d=\Omega(n)$ is required.
    If these channel gates are quantum gates, then there exists a quantum unitary circuit with $d=O(1)$ which transport between the mixed states $\mathrm{diag}(Q)$ and $\mathrm{diag}(P)$. 
    All gates in these circuit admits the local reversibility property.
    }
    \label{fig:1}
\end{figure}

We begin by formalizing the geometric locality constraint used to compare quantum and classical transport.
Suppose a circuit $\mathcal C$ on a graph, we assume it consists of constant layers of local channel gates (either classical or quantum).
Let $\mathcal L^-_{\mathcal C}(v)$ denote the backward lightcone of an output site $v$, consisting of all input sites that can influence it, and let $\mathcal L^+_{\mathcal C}(u)$ denote the forward lightcone of an input site $u$, consisting of all output sites it can influence.
We say that $\mathcal C$ is \emph{bidirectionally geometrically $d$-local}, if $\mathcal L^-_{\mathcal C}(v)\subseteq B_d(v)$ and $\mathcal L^+_{\mathcal C}(u)\subseteq B_d(u)$ for every input site $u$ and output site $v$, where $B_d(v)$ denotes the graph-distance-$d$ neighborhood of $v$.
We call a family of circuits geometrically local if $d=O(1)$, independently of the system size.
We also say a transport from $Q$ to $P$ \emph{locally reversible} if it is implemented by a bidirectionally geometrically local circuit in which the action of every local channel gate on the corresponding intermediate state can be undone by another local channel \cite{sang_2025_mixed, hu_2026_learning}. That is for any intermediate state $\rho$ during transport and any local gate $\mathcal{E}$ in the next transport step, there always exists another local gate $\tilde{\mathcal{E}}$ such that $\tilde{\mathcal{E}} \circ \mathcal{E} (\rho) = \rho$.

With this common locality constraint, we now construct two probability distributions $Q_n$ and $P_n$ on an open one-dimensional chain.
This construction form the basis of our analysis in Sec.\,\ref{sec:def_PQ}--\ref{sec:CH}, where we will see there exists locally reversible quantum transport between $Q_n$ and $P_n$, but there is no bidirectionally geometrically local circuit transporting from $Q_n$ to $P_n$ (see Fig.\,\ref{fig:1}).
In Sec.\,\ref{sec:nonlocal-game-transport}, we generalize this construction to more general graphs using nonlocal games.

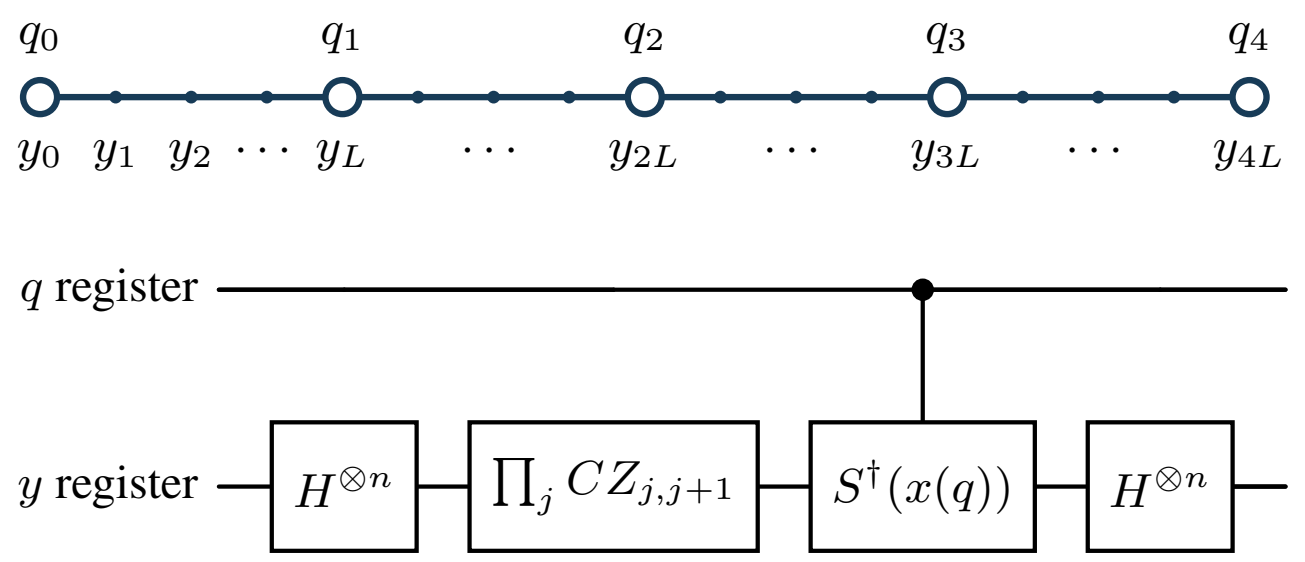
\begin{figure}[t]
\centering
\begin{tikzpicture}[x=1cm,y=1cm]
  \draw[deepblue,very thick] (0,0)--(8,0);
  \foreach \x in {0,2,4,6,8}{
    \fill[white] (\x,0) circle (3.2pt);
    \draw[deepblue,very thick] (\x,0) circle (3.2pt);
  }
  \foreach \x in {0.5,1,1.5,2.5,3,3.5,4.5,5,5.5,6.5,7,7.5}{
    \fill[deepblue] (\x,0) circle (1.2pt);
  }
  \node[above=5pt] at (0,0) {$q_0$};
  \node[above=5pt] at (2,0) {$q_1$};
  \node[above=5pt] at (4,0) {$q_2$};
  \node[above=5pt] at (6,0) {$q_3$};
  \node[above=5pt] at (8,0) {$q_4$};
  \node[below=5pt] at (0.0,0) {$y_0$};
  \node[below=5pt] at (0.5,0) {$y_1$};
  \node[below=5pt] at (1.0,0) {$y_2$};
  \node[below=5pt] at (1.5,0) {$\cdots$};
  \node[below=5pt] at (2.0,0) {$y_{L}$};
  \node[below=5pt] at (3.0,0) {$\cdots$};
  \node[below=5pt] at (4.0,0) {$y_{2L}$};
  \node[below=5pt] at (5.0,0) {$\cdots$};
  \node[below=5pt] at (6.0,0) {$y_{3L}$};
  \node[below=5pt] at (7.0,0) {$\cdots$};
  \node[below=5pt] at (8.0,0) {$y_{4L}$};
\end{tikzpicture}

\vspace{4mm}

\begin{quantikz}[row sep=0.9cm, column sep=0.35cm]
\lstick{$q$ register}
  & \qw & \qw & \ctrl{1} & \qw & \qw \\
\lstick{$y$ register}
  & \gate{H^{\otimes n}}
  & \gate{\prod_j CZ_{j,j+1}}
  & \gate{S^\dagger (x(q))}
  & \gate{H^{\otimes n}}
  & \qw
\end{quantikz}

\vspace{2mm}

\caption{
    Upper panel: One data (qu)bit $y_j$ per path vertex. At each of the five marked sites $p_a$, we attach one question (qu)bit $q_a$.
    Lower panel: Schematic of quantum circuit $U_n$ for the open-chain construction. The $y$ register is acted on by an initial Hadamard layer, the nearest-neighbor open-chain $CZ$ circuit, a layer of local $S^\dagger$ gates controlled by the question register $q$, and a final Hadamard layer.
}
\label{fig:path}
\end{figure}

Let $L$ be odd and set $n=4L+1$.
We place one \textit{data bit} $y_j$ at each vertex $j=0,1,\ldots,4L$ of an open chain and mark the five equally spaced sites $p_a=aL$, with $a=0,1,2,3,4$.
At each marked site $p_a$ we attach an additional local classical bit $q_a\in\{0,1\}$, called \textit{question bit} (see Fig.\,\ref{fig:path}).
We write $q=(q_0,\ldots,q_4)$ and take these five bits to be uniformly distributed.
We remark that the $q_a$'s are part of the one-dimensional system itself rather than externally supplied controls. Thus, eventually we will analyze the distributions defined on these $n+5$ bits.

We embed the 5-bit string $q$ into another binary string $x(q)\in \{0, 1\}^n$ by taking $x_{p_a}(q)=q_a$ and $x_j(q)=0$ elsewhere.
Let $A_\mathrm{adj}$ be the adjacency matrix of the open path and define $M_q=A_\mathrm{adj}+\operatorname{diag}[x(q)]$ over $\{0,1\}$.
The kernel equation $M_q u=0$ is equivalent to a second-order recurrence along the chain, therefore $\dim\ker M_q\leq 1$ (see Appendix \ref{sec:1d}).
Whenever the kernel is nontrivial, we denote its unique nonzero vector by $u_q$.
For $u\in\ker M_q$ (whenever $M_q$ has nontrivial kernel or not), define
\begin{equation}
    \lambda_q(u)
    =
    \sum_{j=0}^{n-2}u_ju_{j+1}
    +
    \frac{1}{2}\sum_{j=0}^{n-1}x_j(q)u_j
    \pmod 2,
    \label{eq:lambda}
\end{equation}
where the second sum is always even on $\ker M_q$, so $\lambda_q$ is well defined.
As shown in Sec.~\ref{sec:QT}, $\ker M_q$ collects exactly the parity checks $Z^u$ that the quantum circuit preserves, and $\lambda_q(u)$ is the sign that each of them acquires.

The source distribution is $Q_q=\operatorname{Unif}(\operatorname{im}M_q)$.
That is, if $\ker M_q = \{0\}$, $Q_q$ is completely uniform; while if $\ker M_q=\operatorname{span}\{u_q\}$, $Q_q$ is uniform on the homogeneous parity sector $u_q^T y=0$.
We define $P_q$ similarly
\begin{equation}
    P_q(y)=
    \begin{cases}
        \frac{1}{2^{n}},
        & \ker M_q =\{0\},\\[0mm]
        \frac{1}{2^{n-1}}
        \,\mathbf 1\!\left[u_q^Ty=\lambda_q(u_q)\right],
        & \ker M_q\neq\{0\}.
    \end{cases}
    \label{eq:Pq-definition}
\end{equation}
Thus, for each fixed $q$, distributions $Q_q$ and $P_q$ are equally simple and differ only in the parity $\lambda_q(u_q)$ of a single parity check; the entire classical obstruction will come from how this parity depends on $q$.

Including the five question bits, the full distributions are $Q_n(q,y)=2^{-5}Q_q(y)$ and $P_n(q,y)=2^{-5}P_q(y)$.
An important feature of the construction is that $Q_n$ itself is locally generated.
For fixed $q$, draw independent unbiased bits $r\in \{0, 1\}^n$ and set $z=M_qr$.
Because $z_j=r_{j-1}+x_j(q)r_j+r_{j+1}$, with the absent boundary terms omitted, this is a radius-one classical sampler.
Thus the source distribution is already easy to prepare locally; the separation below concerns the locality required to transform it into $P_n$.

\section{Quantum transport}
\label{sec:QT}

We now show that $Q_n$ and $P_n$ are connected exactly and reversibly by a constant-depth geometrically local quantum circuit.

For fixed $q$, define a circuit acting on $y$ register (see Fig.\,\ref{fig:path}):
\begin{equation}
    C_q
    =
    H^{\otimes n}
    \left[
    \bigotimes_{j=0}^{n-1}
    \left(S_j^\dagger\right)^{x_j(q)}
    \right]
    \left[
    \prod_{j=0}^{n-2}CZ_{j,j+1}
    \right]
    H^{\otimes n},
    \label{eq:Cq}
\end{equation}
where $S=\operatorname{diag}(1,i)$.
The nearest-neighbor controlled-$Z$ gates can be arranged in two layers, while all Hadamard and phase gates are on-site.
At each marked site $p_a$, the factor $(S_{p_a}^\dagger)^{q_a}$ is implemented by an on-site controlled-$S^\dagger$ gate between the local control bit and the data qubit at the same site.
Hence the full circuit has depth and geometric range independent of $n$ and requires no nonlocal access to $q$.
Applied to $|0^n\rangle$ and followed by a computational-basis measurement, $C_q$ measures the one-dimensional cluster state in the $X$ basis at every site, except at marked sites with $q_a=1$, where it measures $Y$; the question bits thus select measurement bases, exactly as in a nonlocal game.

For a binary vector $u \in \{0, 1\}^n$, write $Z^u=\prod_j Z_j^{u_j}$.
The key property of the circuit is that every parity operator selected by the kernel of $M_q$ is preserved up to the phase defined in Eq.~\eqref{eq:lambda}.
In fact, for all $u\in\ker M_q$ we have $C_q Z^u C_q^\dagger = (-1)^{\lambda_q(u)}Z^u$ and therefore $C_q \operatorname{diag}(Q_q) C_q^\dagger = \operatorname{diag}(P_q)$; see a direct Clifford calculation given in the Appendix \ref{sec:QC_TD}.
Indeed, when $\ker M_q\neq\{0\}$ we have $\operatorname{diag}(Q_q)=2^{-n}(I+Z^{u_q})$ and $\operatorname{diag}(P_q)=2^{-n}\big(I+(-1)^{\lambda_q(u_q)}Z^{u_q}\big)$, while both equal $2^{-n}I$ otherwise.

Let $\rho_{Q_n}$ and $\rho_{P_n}$ denote the diagonal density matrices associated with the full joint distributions, including the question register.
The controlled circuit $U_n=\sum_q |q\rangle\langle q|\otimes C_q$ is implemented locally by the on-site control gates, and implies
\begin{equation}
    U_n\rho_{Q_n}U_n^\dagger=\rho_{P_n},
    \label{eq:bidirectional-transport}
\end{equation}
Thus the two classical distributions are exactly and bidirectionally connected, via $U_n$ and $U_n^\dagger$, by finite-depth geometrically local quantum dynamics.
Although both endpoint states are diagonal in the computational basis, the intermediate states need not remain classical.

\section{Classical hardness}
\label{sec:CH}

We now turn to classical transport by studying the classical depth complexity of $P_n$.
Because $Q_n$ itself has a radius-one classical sampler (Sec.~\ref{sec:def_PQ}), any local classical transport from $Q_n$ to $P_n$ would yield a local classical generator of $P_n$; we therefore prove the stronger statement that $P_n$ cannot be generated locally from product randomness.
We note that our classical transport acts on the full input $(q,y)$, namely all five question bits and all $n$ open chain bits.

We start from an arbitrary product distribution $\Pi$, with no restriction on the number of classical random variables, and apply a possibly stochastic classical circuit whose input-output lightcones have bidirectional geometric radius at most $d$.
We will show that no such circuit can accurately generate $P_n$ when $d$ is sublinear in the system size.

For each marked site $p_a$, define the enlarged neighborhood $N_a:=B_{2d}(p_a)$\xg{.}
If an output random bit $Y_j$ lies outside $N_a$, then its distance from the random output question bit $Q_a$ is greater than $2d$.
Since both backward lightcones have radius at most $d$, their backward lightcones are disjoint.
Because the source distribution is a product distribution, $Y_j$ and $Q_a$ are therefore independent, as long as $4d<L$.
Moreover, when $4d<L$ the neighborhoods $N_a$ are pairwise disjoint, so an output inside $N_a$ can be correlated with $Q_a$ but with no other question bit.

We call a deterministic function $f:q\mapsto y$ \textit{locally controlled} if, for every output site $j$ and the non-overlapping neighborhoods $\{N_a\}$,
\begin{enumerate}
    \item[(i)] if $j\in N_a$, then $f_j(q)$ may depend on $q_a$ but is independent of all $q_b$ with $b\neq a$;
    \item[(ii)] and if $j\notin\bigcup_a N_a$, then $f_j(q)$ is independent of the entire string $q$.
\end{enumerate}
Thus an output outside the five enlarged neighborhoods has a single response value for all questions, whereas an output inside $N_a$ has at most two response values, distinguished only by the local bit $q_a$.
Locally controlled functions are precisely the classical strategies of the associated nonlocal game: each output site acts as a player who sees at most one question bit.

\subsection{All-versus-nothing nonlocal-game obstruction}
\label{sec:NLG}

We first give an important fact, \textit{all-versus-nothing} (AVN) obstruction, before we dive into the classical lower bound.
Particularly, we will identify several question string $q$ for which $\ker M_q$ are all nontrivial.
For each such $q$, Eq.~(\ref{eq:Pq-definition}) requires every output $y$ in the support of $P_q$ to satisfy an affine parity equation.
The key observation is that these parity equations cannot be satisfied simultaneously by any locally controlled assignment $q \mapsto y$.
This gives a nonlocal-game obstruction.
This deterministic contradiction will be the main ingredient of Sec.~\ref{sec:hardness_proof}.

To be more specific, we will focus on the six question strings
$q^{(1)},\ldots,q^{(6)}$ given in Table~\ref{tab:hard-questions}.
For each of them, $M_{q^{(\ell)}}$ has a one-dimensional kernel.
Every output $y^{(\ell)}$ in the support of $P_{q^{(\ell)}}$ must therefore satisfy
\begin{equation}
    \bigoplus_{j=0}^{n-1} u_j^{(\ell)}y_j^{(\ell)}
    =
    \lambda^{(\ell)}.
    \label{eq:hard-parity}
\end{equation}
Let $u^{(\ell)}$ denote its nonzero kernel vector and let $\lambda^{(\ell)}=\lambda_{q^{(\ell)}}(u^{(\ell)})$.
The supports of $\{u^{(\ell)}\}$ and values of $\lambda^{(\ell)}$ are also listed in Table~\ref{tab:hard-questions}.

\begin{table}[t]
    \centering
        \begin{tabular}{c|c|c}
        $q$ & $\operatorname{supp}(u_q)$ & $\lambda_q(u_q)$\\
        \hline
        $00000$ & $\operatorname{Alt}[0,4L]$ & $0$\\
        $00101$ & $\operatorname{Alt}[0,2L]\cup[2L,4L]$ & $1$\\
        $10001$ & $[0,4L]$ & $1$\\
        $10100$ & $[0,2L]\cup\operatorname{Alt}[2L,4L]$ & $1$\\
        $11011$ & $[0,L]\cup\operatorname{Alt}[L,3L]\cup[3L,4L]$ & $0$\\
        $11111$ & $[0,L]\cup\operatorname{Alt}[L,3L]\cup[3L,4L]$ & $0$
    \end{tabular}
    \caption{The six affine parity constraints used in the classical lower bound. For any $a,b$ with the same parity, we define $\operatorname{Alt}[a,b]=\{a,a+2,\ldots,b\}$.}
    \label{tab:hard-questions}
\end{table}

The explicit patterns in Table~\ref{tab:hard-questions} obey two simple cancellation rules.
First, at every site $j$, the six kernel vectors have even total parity $\bigoplus_{\ell=1}^{6}u_j^{(\ell)}=0$.
Second, for any $j\in N_a$, they continue to have even parity after restricting to questions with a fixed local value $q_a=b \in\{0,1\}$
\begin{equation}
    \bigoplus_{\ell:q_a^{(\ell)}=b}u_j^{(\ell)}=0.
    \label{eq:local-cancellations}
\end{equation}
These identities follow directly from the kernel patterns in Table~\ref{tab:hard-questions}; their explicit verification is given in the Appendix \ref{sec:six_hard}.

Now suppose that $q\mapsto y(q)$ is locally controlled, and write $y^{(\ell)}=y(q^{(\ell)})$.
For $j\notin\bigcup_a N_a$, the bit $y_j^{(\ell)}$ is the same for all six questions, so its total contribution to the XOR of Eq.~(\ref{eq:hard-parity}) vanishes by the first cancellation rule.
For $j\in N_a$, the response $y_j^{(\ell)}$ depends only on $q_a^{(\ell)}$.
Grouping the six questions according to $q_a^{(\ell)}=0,1$, the second cancellation rule in Eq.~(\ref{eq:local-cancellations}) again makes the total contribution vanish.

Consequently, bidirectional locality forces the XOR of the six left-hand sides of Eq.~(\ref{eq:hard-parity}) to be zero.
However, Table~\ref{tab:hard-questions} gives
\begin{equation}
    \bigoplus_{\ell=1}^{6}\lambda^{(\ell)}=1,
    \quad\text{whereas}\quad
    \bigoplus_{\ell=1}^{6}
    \left(
    \bigoplus_{j=0}^{n-1}u_j^{(\ell)}y_j^{(\ell)}
    \right)
    =0.
    \label{eq:avn-contradiction}
\end{equation}
The six affine parity equations therefore cannot all be satisfied by any locally controlled response.
Equivalently, every such deterministic response must fail on at least one of the six questions.

\subsection{Arbitrary product sources and stochastic channels}
\label{sec:hardness_proof}

We now promote the nonlocal-game obstruction of Sec.~\ref{sec:NLG} to a lower bound for general classical sampling from product distributions.
We begin with deterministic circuits.

\begin{theorem}[Deterministic classical hardness]
    \label{thm:deterministic-classical-hardness}
    Suppose a random input $R \sim \Pi$.
    Let $F:R\mapsto(Q,Y)$ be a deterministic classical map with bidirectional geometric radius at most $d$, where $Q=(Q_0,\ldots,Q_4)$ denotes the five output question bits and $Y=(Y_0,\ldots,Y_{n-1})$ the remaining output bits.
    If $4d<L$, then
    \begin{equation}
        d_{\mathrm{TV}}\!\left(F(\Pi),P_n\right)\geq\frac{1}{64}.
        \label{eq:deterministic-classical-hardness}
    \end{equation}
\end{theorem}

\begin{proof}
    The key step is to couple the conditional output distributions for different values of the output question string.
    Using the product structure of $\Pi$ and bidirectional $d$-locality, one can construct a common random seed $\omega$ such that, for every question string $q$, the random output $y(\omega,q)$ obeys the conditional law $\operatorname{Prob}(Y\mid Q=q)$.
    The explicit construction of $\omega$ is given in the Appendix \ref{sec:proof_classical_hardness}.
    Crucially, for every fixed realization of $\omega$, the deterministic map $q\mapsto y(\omega,q)$ is a locally controlled function in the sense defined above.
    
    Let $D=F(\Pi)$ and $\varepsilon=d_{\mathrm{TV}}(D,P_n)$.
    If $\varepsilon\ge 1/32$, the theorem is immediate, so we assume $\varepsilon<1/32$.
    Since the $Q$-marginal of $P_n$ is uniform, every question string has probability
    $\mu(q):=D(Q=q)\ge 1/32 - \varepsilon$.
    
    For each hard question $q^{(\ell)}$, let $\eta_\ell$ denote the conditional probability that the output violates Eq.~\eqref{eq:hard-parity}, conditioned on $Q=q^{(\ell)}$.
    For every fixed seed $\omega$, the map $q\mapsto y(\omega,q)$ is locally controlled.
    By Sec.~\ref{sec:NLG}, it must therefore violate at least one of the six affine parity equations.
    Averaging this pointwise statement over $\omega$ gives $\sum_{\ell=1}^6 \eta_\ell \ge 1$.
    
    Let $\mathrm{Win}=\{(q,y):P_q(y)>0\}$ denote the support relation of the target distribution.
    Since $P_n(\mathrm{Win}^c)=0$, total-variation distance implies $D(\mathrm{Win}^c)\le\varepsilon$.
    On the other hand, failure of the parity equation for any hard question places the output outside $\mathrm{Win}$, and hence
    \begin{equation}
        \varepsilon
        \ge
        D(\mathrm{Win}^c)
        \ge
        \sum_{\ell=1}^6
        \mu(q^{(\ell)})\eta_\ell
        \ge
        \frac{1}{32}-\varepsilon.
    \end{equation}
    Therefore $\varepsilon\ge 1/64$, proving the theorem.
\end{proof}

An arbitrary circuit of local stochastic channels reduces to the same setting by exposing the independent private randomness of every stochastic gate and treating it as additional local product-source randomness.
This does not enlarge the geometric lightcones, so the lower bound applies unchanged.
Combined with the radius-one sampler for $Q_n$, this yields our main classical result for both generation and transport:
\begin{theorem}[Stochastic classical hardness]
    \label{thm:classical-sampling-hardness}
    Let $\Pi$ be an arbitrary product distribution, and let $\mathcal{C}$ be any possibly stochastic classical circuit with bidirectional geometric radius at most $d$.
    If $4(d+1)<L$, then
    \begin{equation}
        d_{\mathrm{TV}}\!\left(\mathcal{C}(\Pi),P_n\right)\geq \frac{1}{64}.
        \label{eq:classical-sampling-hardness}
    \end{equation}
    In particular, the same bound holds for $\mathcal{C}(Q_n)$ for any such circuit $\mathcal{C}$ of radius at most $d$, since composing $\mathcal{C}$ with the radius-one sampler of $Q_n$ yields a circuit of radius at most $d$ acting on product randomness.
\end{theorem}

\section{From nonlocal games to transport advantage}
\label{sec:nonlocal-game-transport}

We now present a general construction that transforms suitable nonlocal games into quantum advantages in transporting classical probability distributions. Consider a family of finite $k$-dimensional lattices $\Lambda_L$ with $n$ data qubits and a constant number of marked sites $p_1,\ldots,p_m$, whose pairwise distances diverge with the system size $L$. At each marked site $p_a$, we place a local question variable $q_a$, drawn independently from a fixed distribution $\pi_a$. We write $q=(q_1,\ldots,q_m)$ and $\pi(q)=\prod_a\pi_a(q_a)$. For each question string $q$, let $C_q$ be a unitary acting on the data register which will be used to produce quantum nonlocal game later, and assume that the full question-controlled unitary $U=\sum_q|q\rangle\langle q|\otimes C_q$ admits a geometrically local implementation with lightcone radius $R=O(1)$, independent of $L$.

We next construct two classical distributions $Q$ and $P$ from $\{C_q\}$ which leads to a quantum-classical separation of transport circuit depth.
Our construction requires only that a certain nonlocal game induced by $C_q$ satisfies a suitable geometric AVN obstruction, which we will formulate and explain in detail below.

Since classical probability distributions correspond to diagonal density matrices, we seek distributions that remain diagonal after conjugation by the quantum circuit $C_q$.
As any diagonal density matrix can be expanded in the Pauli-$Z$ basis, this motivates us to identify, for each question $q$, the binary strings $u$ whose corresponding Pauli operators $Z^u$ remain diagonal under this transformation:
\begin{equation}
    K_q:=\left\{u\in\{0, 1\}^n:C_qZ^uC_q^\dagger\text{ is diagonal}\right\}.
    \label{eq:game-Kq}
\end{equation}
Since $Z^{u+v}=Z^uZ^v$ and diagonal operators are closed under multiplication, $K_q$ is a linear subspace of $\{0, 1\}^n$.
Let $\chi=\dim K_q$ and choose a basis $u^{(1)},\ldots,u^{(\chi)}$ of $K_q$, and we will see that the set of $q$ whose $\chi > 0$ must be non-empty as long as the AVN obstruction holds.
Define $D_{q,\ell} := C_q Z^{u^{(\ell)}} C_q^\dagger$. Each $D_{q,\ell}$ is a diagonal Hermitian unitary with eigenvalues $\pm1$, so there exists a Boolean function $h_{q,\ell}:\{0,1\}^n\to\{0,1\}$ such that $D_{q,\ell}|y\rangle=(-1)^{h_{q,\ell}(y)}|y\rangle$. We define the \textit{winning set} of the associated nonlocal game as
\begin{equation}
    W_q:=\left\{y\in\{0,1\}^n:h_{q,\ell}(y)=0,\ \forall\ell\right\}.
    \label{eq:game-winning-set}
\end{equation}
The source and target distributions are then defined by $Q_q:=\operatorname{Unif}(K_q^\perp)$ and $P_q:=\operatorname{Unif}(W_q)$.
Here $K_q^\perp$ denotes the orthogonal complement over $\{0, 1\}$. The constraints defining $P_q$ are generalized parity equations, whose Boolean functions need not be linear. Their common solution set is guaranteed to be nonempty and contains exactly $2^{n-\chi}$ elements, as shown in Appendix \ref{app:general-nonlocal-game}. The full joint distributions are $Q(q,z)=\pi(q)Q_q(z)$ and $P(q,y)=\pi(q)P_q(y)$.

Our construction satisfies two generic properties.
The first important property is that $Q$ can be sampled exactly by a geometrically local classical circuit. To construct such a sampler, consider the Heisenberg-evolved operators $O_{q,j}:=C_q^\dagger Z_j C_q$. It is local because $C_q$ is bidirectionally local. 
Using the Hermitian Pauli notation $\mathsf P(a,b)=i^{a^Tb}X^aZ^b$, we expand $O_{q,j}=\sum_{a,b}c_{q,j}(a,b)\mathsf P(a,b)$ and define $\mathcal A_{q,j}:=\{a:\exists b,\ c_{q,j}(a,b)\neq0\}$. For every site $j$ and every binary vector $a$ supported within $B_R(j)$, introduce an independent unbiased random bit $r_{j,a}$ and the coefficient $g_{q,j,a}:=\mathbf 1[a\in\mathcal A_{q,j}]$. The classical sampler is then
\begin{equation}
z_i=\bigoplus_{j,a}g_{q,j,a}\,r_{j,a}\,a_i.
\label{eq:game-local-sampler}
\end{equation}
Equivalently, writing $(G_q)_{i,(j,a)}=g_{q,j,a}a_i$, we have $z=G_qr$. The key identity is $\operatorname{im}G_q=K_q^\perp$, so independent unbiased randomness generates precisely the uniform distribution $Q_q$. Moreover, the locality of $U$ implies that every contributing vector $a$ is supported within $B_R(j)$ and that $g_{q,j,a}$ depends only on nearby question variables. Consequently, the sampler has bidirectional geometric radius at most $2R=O(1)$. We prove both the sampling identity and its locality in Appendix \ref{app:general-nonlocal-game}.

The second property of our construction is that the distributions $Q$ and $P$ are connected exactly by unitary quantum dynamics.
Since $Q_q$ is uniform on $K_q^\perp$, its diagonal density matrix has the Fourier expansion $\operatorname{diag}(Q_q)=2^{-n}\sum_{u\in K_q}Z^u$.
Meanwhile, the defining constraints of $P_q$ give
\begin{equation}
    \operatorname{diag}(P_q) =2^{\chi-n}\prod_{a=1}^{\chi}\frac{I+D_{q,a}}{2} = 2^{-n}\!\!\sum_{u\in K_q}\!C_qZ^uC_q^\dagger.
    \label{eq:game-P-Fourier}
\end{equation}
The second equality follows because the basis vectors $u^{(a)}$ generate $K_q$ and $D_{q,\ell}=C_qZ^{u^{(\ell)}}C_q^\dagger$. Comparing $\operatorname{diag}(Q_q)$ and $\operatorname{diag}(P_q)$ immediately yields $C_q\operatorname{diag}(Q_q)C_q^\dagger=\operatorname{diag}(P_q)$ (see Appendix \ref{app:general-nonlocal-game}).
Including the question register, we obtain $U\operatorname{diag}(Q)U^\dagger=\operatorname{diag}(P)$.
The transport is exact and reversible, with no measurement, dephasing, or discarding of auxiliary systems.

Finally, it remains to impose a classical obstruction through the nonlocal game.
Here comes to the core assumption we need, that is, the winning sets $W_q$ satisfy a geometric AVN nonlocal-game obstruction.
To be more specific, suppose a finite set of hard questions $\mathcal H$, with $p_*:=\min_{q\in\mathcal H}\pi(q)>0$ independent of $L$.
Our geometric AVN obstruction says that for every fixed $d$ and sufficiently large $L$, no locally controlled response $f:q\mapsto y$ satisfies $f(q)\in W_q$ for every $q\in\mathcal H$.
In particular, this obstruction requires $\dim K_q>0$ for at least some hard questions $q\in\mathcal H$, since otherwise $W_q=\{0,1\}^n$ for every $q\in\mathcal H$ and any deterministic classical response would satisfy all winning conditions.

Under these conditions, the product-source coupling argument of Sec.~\ref{sec:hardness_proof} gives the following general result.
\begin{theorem}[Nonlocal-game transport theorem]
    For any fixed classical locality radius $d$ and sufficiently large $L$ such that $\min_{a \neq b}\mathrm{dist}(p_a, p_b) > 4d$, every possibly stochastic bidirectionally $d$-local classical circuit $\mathcal C$, acting on an arbitrary product distribution $\Pi$, satisfies
    \begin{equation}
        d_{\mathrm{TV}}\!\left(\mathcal C(\Pi),P\right) \geq \frac{p_*}{2}.
    \label{eq:game-classical-hardness}
    \end{equation}
\end{theorem}

Finally, our one-dimensional open-chain construction is a special case of this theorem. The corresponding circuit $C_q$ is Clifford and obeys $C_qZ^uC_q^\dagger=(-1)^{\lambda_q(u)}\mathsf P(M_qu,u)$ for $u \in \ker M_q$, where $M_q$ is the matrix defined in Sec.~\ref{sec:def_PQ}. Consequently, $K_q=\ker M_q$ and $K_q^\perp=\operatorname{im}M_q$, recovering the original source distribution.
For $u\in K_q$, the generalized parity function reduces to $h_{q,u}(y)=u^Ty+\lambda_q(u)$, reproducing the affine parity constraints defining $P_q$.
The six hard questions of Sec.~\ref{sec:CH} provide the required geometric nonlocal-game obstruction, with $p_*=2^{-5}$, recovering the total-variation lower bound $1/64$.

\section{Discussion}

We have established a general framework for constructing quantum-classical separations in the local transport of probability distributions using nonlocal games.
Our construction yields distributions that are connected exactly and reversibly by constant-depth local quantum circuits, while classical stochastic transport is obstructed by the geometric constraints of the associated nonlocal games.
We demonstrate this separation explicitly on a one-dimensional chain, where classical transport with total-variation error below $1/64$ requires a geometric range $\Omega(n)$, and extend the framework to more general geometries.
Since the source distributions are themselves locally classically samplable, our results also establish an advantage in distribution generation.
More broadly, they reveal a fundamental distinction between classical and quantum local reversibility, even when both endpoint states are entirely classical.


The present construction is not expected to define a stable phase under generic local perturbations \cite{sang_2025_mixed}.
Both $P_n$ and $Q_n$ contain system-spanning parity constraints and therefore exhibit long-range conditional correlations.
This structure is reminiscent of the long-range conditional information appearing in strong-to-weak spontaneous symmetry-breaking settings \cite{lee_2025_symmetry, lee_2023_quantum, lessa_2025_strong, kuno_2025_strong}, and it is also what enables the all-versus-nothing obstruction.
At the same time, these global parity constraints are sensitive to suitable local noise and lie outside the finite-Markov-length regime used to characterize locally stable classical and mixed-state phases.

A natural next question is whether a quantum transport advantage can survive when the endpoint distributions are required to have finite correlation length or finite Markov length.
Such a construction would turn the present local-transport separation into a robust distinction between classical and quantum notions of phase equivalence.
We note that such transport advantage cannot be achieved if $P$ is in the same quantum phase of product states, namely the quantum \textit{trivial phase}.
This is because any distribution in the quantum trivial phase possesses the so-called quantum \textit{local recovery} and \textit{local extension} maps \cite{hu_2026_learning}.
By dephasing these maps, it yields classical local recovery and local extension maps which shows $P$ is also classically easy to learn and generate. 
Nevertheless, determining whether an transport obstruction with finite Markov length exists or not for generic nontrivial phase can lead to a deeper understanding of the quantum and classical phase separation.


\section*{Acknowledgement}

F.\,H.\,acknowledges support from the QuEra Quantum Innovation Postdoctoral Fellowship.
X.\,G.\,acknowledges support from NSF-CCF-2534807, NSF PFC grant No.\,PHYS 2317149 and DOE QSA.
The idea of constructing quantum--classical transport separations from nonlocal games originated with the authors.
Part of the proof techniques were developed with the assistance of GPT-5.6, GPT-6 Astra, and Claude Opus 5.5.
The authors subsequently audited and rewrote these parts of proofs and takes full responsibility for the correctness of all results.

\bibliography{ref}

\clearpage

\appendix

\begin{widetext}
    \include{appendix}

\end{widetext}

\end{document}

%% file: appendix.tex
\begin{center}
    {\large \textbf{Supplementary Materials}}\\[1em]
\end{center}

\startcontents[appendices]
\printcontents[appendices]{l}{1}{
    \section*{CONTENTS}
    \setcounter{tocdepth}{2}
    }

\setcounter{page}{1}
\setcounter{equation}{0}
\setcounter{figure}{0}
\counterwithout{equation}{section}
\renewcommand{\tablename}{Table}
\renewcommand{\figurename}{Fig.}
\renewcommand{\thetable}{S\arabic{table}}
\renewcommand{\thefigure}{S\arabic{figure}}
\renewcommand{\theequation}{S\arabic{equation}}
\renewcommand{\thetheorem}{S\arabic{theorem}}

\renewcommand{\theHtable}{S\arabic{table}}
\renewcommand{\theHfigure}{S\arabic{figure}}
\renewcommand{\theHequation}{S\arabic{equation}}
\renewcommand{\theHtheorem}{S\arabic{theorem}}

\renewcommand\thesection{S\arabic{section}}
\renewcommand\thesubsection{\Alph{subsection}}
\renewcommand\thesubsubsection{\arabic{subsubsection}}

\titleformat{\section}
  {\centering\normalfont\bfseries} 
  {\thesection}{1em}{\MakeTextUppercase} 

\titleformat{\subsection}
  {\centering\normalfont\bfseries} 
  {\thesubsection}{1em}{} 

\cftsetindents{section}{0em}{3em}
\cftsetindents{subsection}{3em}{3em}

\setlength{\parindent}{0em}

\makeatletter
\let\toc@pre\relax
\let\toc@post\relax
\makeatother

\section{Probability notation and total variation distance}
\label{sec:app-probability-preliminaries}

We collect here the probability notation and elementary properties of total variation distance used throughout the Supplementary Materials. All random variables below take values in finite sets unless stated otherwise.

\subsection{Probability notation}
\label{subsec:app-probability-notation}

For a random variable $X$, we write $\Law(X)$ for its probability law. More generally, $\Law(X,Y)$ denotes the joint law of two jointly distributed random variables $(X,Y)$.

If $D$ is a probability distribution and $E$ is an event, we write $D(E)$ for the probability of $E$ under $D$. Thus, if $D=\Law(Q,Y)$, then $D(Q=q)$ denotes the probability that $Q=q$, while $D((Q,Y)\in E)$ denotes the probability of an event involving the joint variables $(Q,Y)$.

Given $D=\Law(X,Y)$, we denote the marginal law of $X$ by $D_X=\Law(X)$. Explicitly, $D_X(x)=\sum_yD(x,y)$. When there is no ambiguity, we simply write $D(X=x)$ instead of $D_X(x)$.

If $D(Y=y)>0$, the conditional law of $X$ given $Y=y$ is denoted by $\Law(X\mid Y=y)$, or equivalently by $D_{X\mid Y=y}$, with
\begin{equation}
    D_{X\mid Y=y}(x)=\frac{D(X=x,Y=y)}{D(Y=y)}.
\end{equation}
Similarly, for an event $E$, we write
$D(E\mid Y=y)=D(E\cap{Y=y})/D(Y=y)$.


\subsection{Total variation distance}
\label{subsec:app-tv-distance}

For two probability distributions $D$ and $D'$ on the same finite sample space $\Omega$, their total variation distance is
\begin{equation}
\TV(D,D')
    :=\frac12\sum_{\omega\in\Omega}|D(\omega)-D'(\omega)|
    =\max_{E\subseteq\Omega}|D(E)-D'(E)|.
    \label{eq:app-tv-definition}
\end{equation}
In particular, every event $E$ satisfies
\begin{equation}
    |D(E)-D'(E)|\leq\TV(D,D')
    \label{eq:tv-mon-margin}
\end{equation}

Total variation distance is non-increasing under a classical stochastic map. If $\Phi$ is any classical channel, then
\begin{equation}
    \TV\!\left(\Phi(D),\Phi(D')\right)
    \leq \TV(D,D').
    \label{eq:app-tv-data-processing}
\end{equation}
Marginalization is a special case. Thus, if $D$ and $D'$ are joint distributions of $(X,Y)$, then $\TV(D_X,D'_X)\leq\TV(D,D')$. In particular, for every value $x$,
\begin{equation}
    |D(X=x)-D'(X=x)|\leq\TV(D_X,D_X')\leq\TV(D,D').
    \label{eq:tv-mon-x}
\end{equation}


\section{1D open-boundary chain and local generative models}

\subsection{1D open-boundary chain}
\label{sec:1d}
Fix an odd integer $L$ and set
\begin{equation}
    n=4L+1.
\end{equation}
Place one data qubit at every vertex of the open path $0,1,\ldots,4L$. 

Mark the five sites
\begin{equation}
    p_a=aL,\qquad a=0,1,2,3,4.
\end{equation}
The \textit{question bitstring} is a uniformly random string
\begin{equation}
    q=(q_0,q_1,q_2,q_3,q_4)\in\bits^5.
\end{equation}
Embed it into a length-$n$ basis string $x(q)$ by
\begin{equation}
    x_j(q)=
    \begin{cases}
    q_a,&j=p_a,\\
    0,&j\notin\{p_0,\ldots,p_4\}.
    \end{cases}
\end{equation}

Let $A_\mathrm{adj}\in\F_2^{n\times n}$ be the adjacency matrix of the path and define
\begin{equation}
    M_q:=A_\mathrm{adj}+\diag(x(q)) \qquad\text{over }\F_2.
\end{equation}
The relevant recurrence is
\begin{equation}
    u_{j+1}=x_j(q)u_j+u_{j-1},
    \qquad u_{-1}=u_n=0,
    \label{eq:recurrence}
\end{equation}
with all arithmetic modulo two.

\begin{lemma}[At most one parity constraint]
    For every $q$, $\dim\ker M_q\le 1$.
\end{lemma}
\begin{proof}
    If $u_0=0$, the our convention of $u_{-1}=0$ will enforce $u=0$. Thus, all nontrivial solution to $M_q$ must have $u_0=1$.
    Once $u_0=1$ is fixed, the boundary value $u_{-1}=0$ and the recurrence Eq.\,\eqref{eq:recurrence} determine every subsequent component.  Hence there is at most one nonzero solution.
\end{proof}

\subsection{Classical source, outputs, and locality}
\label{subsec:app-classical-source-locality}

We allow an arbitrary product classical source
\begin{equation}
R=(R_1,\ldots,R_M)\sim\Pi,
\qquad
\Pi=\bigotimes_{m=1}^{M}\pi_m,
\end{equation}
where each $R_m$ is assigned to a site of the one-dimensional chain. The number $M$ may exceed $n$, and arbitrarily many variables may occupy the same site. For a set of sites $A$, we write $R|_A$ for all source variables located in $A$.

The output is
\begin{equation}
(Q,Y)\in\bits^5\times\bits^n.
\end{equation}
Every site $j$ outputs one bit $Y_j$, while each marked site $p_a=aL$ additionally outputs $Q_a$. Distances are always measured along the one-dimensional chain; variables on the same site have distance zero.

We also denote $B_d(p)$ as the $d$-neighborhood of site $p$:
\begin{equation}
    \Ball{d}{p} =\{v\in\{0,\ldots,n-1\}:|v-p|\le d\}.
\end{equation}

\begin{definition}[Deterministic classical transport]
A deterministic \textit{bidirectionally geometrically $d$-local transport} from $\Pi$ is a map
\begin{equation}
    F:R\mapsto(Q,Y).
\end{equation}
It has bidirectional geometric radius at most $d$ if every output depends only on input variables within distance $d$, and every input variable can affect only outputs within distance $d$. Its output law is denoted by $F(\Pi)$.
\end{definition}

\begin{definition}[Noisy classical transport]
A noisy classical transport from $\Pi$ is a circuit of local stochastic channels producing $(Q,Y)$. Its geometric radius is defined by the input-output lightcones of the circuit wiring, and its output law is denoted by $\mathcal C(\Pi)$.
\end{definition}

Each local stochastic gate can equivalently be implemented by adjoining an independent ancillary random variable, applying a deterministic local function, and discarding the ancilla. Therefore, after exposing all gate randomness, any noisy local circuit becomes a deterministic local map acting on an enlarged product source, without enlarging its geometric lightcones. We use this equivalence in Sec.~\ref{sec:noisy}.

\section{Quantum circuit and target distribution}
\label{sec:QC_TD}

We first define the quantum circuit acting on the full $(Q,Y)$ system. At each marked site $p_a$, the control qubit $Q_a$ acts only on the data qubit at the same site through the controlled-$S^\dagger$ gate
\begin{equation}
    \mathrm{CS}_{a}^{\dagger}
    :=
    \ket{0}_{Q_a}\!\bra{0}_{Q_a}\otimes I_{Y_{p_a}}
    +
    \ket{1}_{Q_a}\!\bra{1}_{Q_a}\otimes S_{Y_{p_a}}^\dagger .
    \label{eq:controlled-S}
\end{equation}
We then define
\begin{equation}
    U_n
    :=
    \left(I_Q\otimes H^{\otimes n}\right)
    \left(\prod_{a=0}^{4}\mathrm{CS}_{a}^{\dagger}\right)
    \left(I_Q\otimes\prod_{j=0}^{n-2}CZ_{j,j+1}\right)
    \left(I_Q\otimes H^{\otimes n}\right).
    \label{eq:full-quantum-circuit}
\end{equation}
All controlled phase gates act within a single site, while the $CZ$ gates act only between neighboring data qubits. Hence $U_n$ is a constant-depth geometrically local circuit and leaves the control register $Q$ unchanged.

For a fixed computational-basis value $q\in\bits^5$ of the control register, the action of $U_n$ reduces to
\begin{equation}
    U_n \left(\ket{q}_Q\otimes\ket{\psi}_Y\right)
    =
    \ket{q}_Q\otimes C_q\ket{\psi}_Y,
\end{equation}
where
\begin{equation}
    C_q
    =
    H^{\otimes n}
    \left(\bigotimes_{j=0}^{n-1}(S_j^\dagger)^{x_j(q)}\right)
    \left(\prod_{j=0}^{n-2}CZ_{j,j+1}\right)
    H^{\otimes n}.
    \label{eq:Cq-apx}
\end{equation}
Equivalently, we write
\begin{equation}
    U_n
    =
    \sum_{q\in\bits^5}
    \ket{q}_Q\!\bra{q}_Q\otimes C_q.
\end{equation}
Read Eq.\,(\ref{eq:Cq-apx}) from right to left, this prepares $\lvert+\rangle^{\otimes n}$, creates the open path-cluster state, rotates every requested $Y$-register measurement to computational-basis readout, and finally applies Hadamards. 
We can treat the setting here be a quanutm non-local game, whose local measurement rule is
\begin{equation}
    x_j=0:\ X_j\text{ measurement},
    \qquad
    x_j=1:\ Y_j\text{ measurement}.
\end{equation}


\begin{lemma}[Kernel phase]
    \label{lem:app-kernel-phase}
    For every $u\in\ker M_q$,
    \begin{equation}
    C_q Z^u C_q^\dagger
    =
    (-1)^{\lambda_q(u)}Z^u,
    \label{eq:app-kernel-phase}
    \end{equation}
    where
    \begin{equation}
    \lambda_q(u)
    :=
    \sum_{j=0}^{n-2}u_j u_{j+1}
    +
    \frac{1}{2}\sum_{j=0}^{n-1}x_j(q)u_j
    \pmod 2.
    \label{eq:app-lambda}
    \end{equation}
    The second sum is even for every $u\in\ker M_q$, so $\lambda_q(u)$ is well defined. 
\end{lemma}
    
\begin{proof}
Conjugating $Z^u$ by the first Hadamard layer gives $X^u$. Under the $CZ$ layer, each $X_j$ acquires a $Z$ on each neighboring site. Reordering the resulting Pauli product into the form $X^uZ^{Au}$ contributes one minus sign for every edge whose two endpoints lie in $\supp(u)$. Hence
\begin{equation}
X^u
\longmapsto
(-1)^{\sum_{j=0}^{n-2}u_ju_{j+1}}
X^uZ^{Au}.
\label{eq:app-CZ-action}
\end{equation}

For $u\in\ker M_q$, we have $Au=x(q)\odot u$. Thus, whenever
$x_j(q)u_j=1$, the local Pauli factor is $X_jZ_j$. Using
$S^\dagger XZS=-iX$, the $S^\dagger$ layer gives
\begin{equation}
X^uZ^{Au}
\longmapsto
(-i)^{N_q(u)}X^u,
\qquad
N_q(u):=\sum_{j=0}^{n-1}x_j(q)u_j.
\label{eq:app-S-action}
\end{equation}
Moreover, the kernel condition implies $N_q(u)=u^T\diag(x(q))u=u^TAu=0\pmod 2$, since $A$ is symmetric with zero diagonal. Hence $N_q(u)$ is even and $(-i)^{N_q(u)}=(-1)^{N_q(u)/2}$.

The final Hadamard layer maps $X^u$ back to $Z^u$, yielding
\begin{equation}
C_qZ^uC_q^\dagger
=
(-1)^{\lambda_q(u)}Z^u,
\end{equation}
with
\begin{equation}
\lambda_q(u)
=
\sum_{j=0}^{n-2}u_ju_{j+1}
+
\frac{1}{2}\sum_{j=0}^{n-1}x_j(q)u_j
\pmod 2.
\end{equation}
\end{proof}

\section{Locally sampled source and exact transport}

Define the source distribution
\begin{equation}
    Q_q:=\Unif(\im M_q).
    \label{eq:Qq}
\end{equation}
It has a radius-one sampler. In fact, draw independent unbiased bits $r\in\F_2^n$ and output
\begin{equation}
    z=M_qr.
    \label{eq:sampler}
\end{equation}
Explicitly,
\begin{equation}
    z_j=r_{j-1}+x_j(q)r_j+r_{j+1},
\end{equation}
with the absent boundary terms omitted.  Every output uses only the local question bit, the local random bit, and its two nearest random neighbors.

Since $M_q$ is symmetric,
\begin{equation}
    \im M_q=(\ker M_q)^\perp.
\end{equation}
Hence, if $\ker M_q=\operatorname{span}\{u_q\}$,
\begin{equation}
    Q_q(z)=2^{-(n-1)}\mathbf 1[u_q^Tz=0],
\end{equation}
and otherwise $Q_q$ is fully uniform.

\begin{lemma}[Fourier expansions for nontrivial sectors]
    \label{lem:app-fourier-nontrivial}
    Suppose $\ker M_q=\operatorname{span}{u_q}$ with $u_q\neq0$. Then
    \begin{equation}
        \operatorname{diag}(Q_q)
        =
        2^{-n}\left(I+Z^{u_q}\right),
        \label{eq:app-Qq-fourier-nontrivial}
    \end{equation}
    and
    \begin{equation}
        \operatorname{diag}(P_q)
        =
        2^{-n}\left(
        I+(-1)^{\lambda_q(u_q)}Z^{u_q}
        \right).
        \label{eq:app-Pq-fourier-nontrivial}
    \end{equation}
\end{lemma}

\begin{proof}
    Since $\ker M_q=\operatorname{span}{u_q}$ and $M_q$ is symmetric,
    $\operatorname{im}M_q=(\ker M_q)^\perp$. Hence
    $Q_q$ is uniform on the hyperplane
    $u_q^Tz=0$. Using
    \begin{equation}
        \mathbf 1{u_q^Tz=0}
        =
        \frac12\left(1+(-1)^{u_q^Tz}\right),
    \end{equation}
    we have
    \begin{equation}
        Q_q(z)
        =
        2^{-n}
        \left(1+(-1)^{u_q^Tz}\right).
    \end{equation}
    Therefore
    \begin{equation}
        \operatorname{diag}(Q_q)
        =
        2^{-n}
        \left(
        I+\sum_z(-1)^{u_q^Tz}\ket{z}\!\bra{z}
        \right)
        =
        2^{-n}\left(I+Z^{u_q}\right).
    \end{equation}
    
    Similarly, $P_q$ is uniform on the affine hyperplane
    $u_q^Ty=\lambda_q(u_q)$. Thus
    \begin{equation}
        P_q(y)
        =
        2^{-n}
        \left(
        1+(-1)^{u_q^Ty+\lambda_q(u_q)}
        \right),
    \end{equation}
    and hence
    \begin{equation}
        \operatorname{diag}(P_q)
        =
        2^{-n}
        \left(
        I+(-1)^{\lambda_q(u_q)}
        \sum_y(-1)^{u_q^Ty}\ket{y}\!\bra{y}
        \right)
        =
        2^{-n}
        \left(
        I+(-1)^{\lambda_q(u_q)}Z^{u_q}
        \right),
    \end{equation}
    which completes the proof.
\end{proof}

Now, we can eventually prove the exact transport:

\begin{theorem}[Exact conditional transport]
    For every question $q$,
    \begin{equation}
        C_q\,\diag(Q_q)\,C_q^\dagger=\diag(P_q).
    \label{eq:exacttransport}
    \end{equation}
\end{theorem}
\begin{proof}
    If $\ker M_q = \mathrm{span} \{u_q\}$, then $Q_q$ is
    \begin{equation}
        \diag(Q_q)=2^{-n}\sum_{u\in\ker M_q}Z^u.
    \end{equation}
    Conjugating term-by-term and using Eq.\,(\ref{eq:app-kernel-phase}) gives
    \begin{equation}
        C_q\diag(Q_q)C_q^\dagger
        =2^{-n}\sum_{u\in\ker M_q}(-1)^{\lambda_q(u)}Z^u = \diag(P_q),
    \end{equation}
    which completes the proof.
\end{proof}

Consider the full state including $Q$-register:
\begin{equation}
    \rho_Q=2^{-5}\sum_q\ketbra q\otimes\diag(Q_q),
    \qquad
    \rho_P=2^{-5}\sum_q\ketbra q\otimes\diag(P_q),
\end{equation}
Recall that $U=\sum_q\ketbra q\otimes C_q$. Then,
\begin{equation}
    U\rho_QU^\dagger=\rho_P,
    \qquad
    U^\dagger\rho_PU=\rho_Q.
\end{equation}
Each local gate is inverted on the same support, although the intermediate states need not remain diagonal.

\section{Six hard questions}
\label{sec:six_hard}

For integers $a\le b$, write $[a,b]=\{a,a+1,\ldots,b\}$; when $a$ and
$b$ have the same parity, write
\begin{equation}
    \Alt[a,b]=\{a,a+2,a+4,\ldots,b\}.
\end{equation}

The support calculation is local.  At a marked site $j=p_a$, we have
$x_j(q)=q_a$, so the recurrence Eq.\,\eqref{eq:recurrence} becomes
\begin{equation}
    u_{j+1}=q_a u_j+u_{j-1}.
    \label{eq:markedtransition}
\end{equation}
We use $u$ in the transition table below in order to distinguish the kernel vector from the basis-choice string $x(q)$.  Once the incoming pair $(u_{j-1},u_j)$ and the question bit $q_a$ are known, Eq.\,\eqref{eq:markedtransition} fixes $u_{j+1}$.

\begin{center}
    \small
    \renewcommand{\arraystretch}{1.16}
    \begin{tabular}{@{}c c c c l@{}}
    \toprule
    $q_a=x_{p_a}(q)$ & $u_{j-1}$ & $u_j$ & $u_{j+1}$
    & Pattern on the unmarked sites after $j$\\
    \midrule
    $0$ & $0$ & $0$ & $0$ & all zeros\\
    $0$ & $0$ & $1$ & $0$ & alternating, with a $1$ at $j$\\
    $0$ & $1$ & $0$ & $1$ & alternating, with a $1$ at $j+1$\\
    $0$ & $1$ & $1$ & $1$ & all ones\\
    \addlinespace[0.2em]
    $1$ & $0$ & $0$ & $0$ & all zeros\\
    $1$ & $0$ & $1$ & $1$ & all ones\\
    $1$ & $1$ & $0$ & $1$ & alternating, with a $1$ at $j+1$\\
    $1$ & $1$ & $1$ & $0$ & alternating, with a $1$ at $j$\\
    \bottomrule
    \end{tabular}
    
    \vspace{0.35em}
    \textit{Local transition rule at a marked site $j=p_a$.}
\end{center}

Indeed, every site strictly between two consecutive marked sites has
$x_k(q)=0$.  There the recurrence reduces to $u_{k+1}=u_{k-1}$.
Consequently, the outgoing pair $(u_j,u_{j+1})$ propagates as one of only
three patterns: all zeros, all ones, or alternating zeros and ones.  The two
alternating rows record the phase of that pattern, namely whether its first
$1$ occurs at $j$ or at $j+1$.  Because the separation $L$ is odd, an
alternating pattern arrives at the next marked site in the opposite phase:
the anchor triple $010$ is followed by $101$, and conversely.

Consider the six question strings
\begin{equation}
00000,\quad00101,\quad10001,\quad10100,\quad11011,\quad11111.
\label{eq:sixq}
\end{equation}
For each one, $M_q$ has a one-dimensional kernel.  To obtain its nonzero
vector, the left boundary condition gives $u_{-1}=0$, and we choose the only
nonzero normalization $u_0=1$.  Starting at $p_0=0$ and repeatedly applying
the local table gives the following triples at the five marked sites:
\begin{center}
\small
\renewcommand{\arraystretch}{1.10}
\begin{tabular}{@{}c c@{}}
\toprule
$q$ & $\bigl(u_{p_a-1}\,u_{p_a}\,u_{p_a+1}\bigr)_{a=0}^{4}$\\
\midrule
$00000$ & $010\;\longrightarrow\;101\;\longrightarrow\;010\;\longrightarrow\;101\;\longrightarrow\;010$\\
$00101$ & $010\;\longrightarrow\;101\;\longrightarrow\;011\;\longrightarrow\;111\;\longrightarrow\;110$\\
$10001$ & $011\;\longrightarrow\;111\;\longrightarrow\;111\;\longrightarrow\;111\;\longrightarrow\;110$\\
$10100$ & $011\;\longrightarrow\;111\;\longrightarrow\;110\;\longrightarrow\;101\;\longrightarrow\;010$\\
$11011$ & $011\;\longrightarrow\;110\;\longrightarrow\;101\;\longrightarrow\;011\;\longrightarrow\;110$\\
$11111$ & $011\;\longrightarrow\;110\;\longrightarrow\;101\;\longrightarrow\;011\;\longrightarrow\;110$\\
\bottomrule
\end{tabular}
\end{center}
The final entry of each chain has $u_{4L+1}=u_n=0$, so it also verifies the
right boundary condition.  Reading $010$ and $101$ as alternating segments,
and $011$, $111$, and $110$ as the entrance to, interior of, and exit from an
all-one segment, respectively, gives the support column of Table~\ref{tab:hard}:
\begin{itemize}
\item $00000$ alternates across the whole path, giving $\Alt[0,4L]$.
\item $00101$ alternates through $2L$ and is then all one, giving
$\Alt[0,2L]\cup[2L,4L]$.
\item $10001$ is all one from the left endpoint to the right endpoint, giving
$[0,4L]$.
\item $10100$ is all one through $2L$ and then alternates, giving
$[0,2L]\cup\Alt[2L,4L]$.
\item Both $11011$ and $11111$ are all one on the two end intervals and
alternate across the middle two intervals, giving
$[0,L]\cup\Alt[L,3L]\cup[3L,4L]$.  The middle question bit does not
distinguish these two rows because $u_{2L}=0$.
\end{itemize}

It remains to compute the last column of Table~\ref{tab:hard}.  Let
\[
E_q:=\sum_{j=0}^{n-2}u_ju_{j+1},
\qquad
N_q:=\sum_{j=0}^{n-1}x_j(q)u_j.
\]
Here $E_q$ counts adjacent pairs of ones, while $N_q$ counts marked sites at
which both $q_a$ and $u_{p_a}$ are one.  The supports just obtained give
\[
\begin{array}{c|cccccc}
q&00000&00101&10001&10100&11011&11111\\
\midrule
E_q&0&2L&4L&2L&2L&2L\\
N_q&0&2&2&2&4&4
\end{array}
\]
and therefore Eq.\,(\ref{eq:app-kernel-phase}), namely $\lambda_q(u_q)=E_q+N_q/2\pmod 2$, gives precisely $0,1,1,1,0,0$.  This proves every entry in Table~\ref{tab:hard} directly from the
local transition rule.

\begin{table}[t]
\centering
\small
\renewcommand{\arraystretch}{1.22}
\begin{tabular}{@{}c l c@{}}
\toprule
$q$ & $\supp(u_{q})$ & $\lambda_q(u_q)$\\
\midrule
$00000$ & $\Alt[0,4L]$ & $0$\\
$00101$ & $\Alt[0,2L]\cup[2L,4L]$ & $1$\\
$10001$ & $[0,4L]$ & $1$\\
$10100$ & $[0,2L]\cup\Alt[2L,4L]$ & $1$\\
$11011$ & $[0,L]\cup\Alt[L,3L]\cup[3L,4L]$ & $0$\\
$11111$ & $[0,L]\cup\Alt[L,3L]\cup[3L,4L]$ & $0$\\
\bottomrule
\end{tabular}
\caption{The six hard affine parity constraints, computed from the local transition rule above.}
\label{tab:hard}
\end{table}

In the next section, we will leverage Table \ref{tab:hard} to turn it into a contradiction, for any classical deterministic transport functions with only local lightcones.

\section{Lower Bound for classical deterministic transport function}

\begin{definition}[Bidirectional geometric radius]
    A deterministic classical transport map has bidirectional geometric radius at most $d$ if every output at site $v$ depends only on input bits within distance $d$ of $v$, and changing an input coordinate at site $u$ can affect only outputs within distance $d$ of $u$.
\end{definition}

Now, we define two important concepts of neighborhoods:
\begin{definition}[Source and enlarged neighborhoods]
    \label{def:AN_neighborhoods}
    For each marked site $p_a$, define the source neighborhood
    \begin{equation}
        A_a:=\Ball{d}{p_a} =\{v\in\{0,\ldots,n-1\}:|v-p_a|\le d\},
    \end{equation}
    and the enlarged neighborhood
    \begin{equation}
        N_a:=\Ball{2d}{p_a}=\{j\in\{0,\ldots,n-1\}:|j-p_a|\le 2d\}.
    \end{equation}
\end{definition}
Clearly $A_a\subseteq N_a$. Since $4d<L$, the five sets $N_0,\ldots,N_4$ are pairwise disjoint; in particular, the five sets $A_0,\ldots,A_4$ are pairwise disjoint. Let
\begin{equation}
    A_{\mathrm{out}}:=\{0,\ldots,n-1\}\setminus\bigcup_{a=0}^4 A_a.
\end{equation}
By geometric locality, $Q_a$ is a deterministic function only of $R|_{A_a}$.

In this section, our main goal is to prove:

\begin{theorem}[Lower bound for local deterministic transport function, Theorem \ref{thm:deterministic-classical-hardness} in main text]
    \label{thm:local deterministic}
    Let $R=(R_1,\ldots,R_{M})\sim\Pi$ be an arbitrary product classical random source lying on the 1D chain.
    Let $(Q,Y)=F(R)$ for a deterministic classical map $F$ of bidirectional geometric radius at most $d$. The bit $Q_a$ is regarded as an output at site $p_a$, while $Y_j$ is an output at site $j$.
    If
    \begin{equation}
        4d<L, \quad \Leftrightarrow \quad d < \frac{L}{4}=\frac{n-1}{16},
    \end{equation}
    then
    \begin{equation}
        \TV\bigl(F(\Pi),P_n\bigr)\ge \frac{1}{64}.
    \end{equation}
\end{theorem}

\subsection{Locally-controlled function}

From now on, for clarification, we denote $q^{(1)},\ldots,q^{(6)}$ be the six questions in the order displayed in Eq.\,\eqref{eq:sixq}, $u^{(\ell)}$ be the corresponding nonzero kernel vector, and $\lambda^{(\ell)}=\lambda_{q^{(\ell)}}(u^{(\ell)})$. 
These $u^{(\ell)}$ satisfy the following properties, stated in Lemma \ref{lem:globalcancel} and Lemma \ref{lem:localcancel}.

\begin{lemma}
    \label{lem:globalcancel}
    For any $j \in [n]$, we have
    \begin{equation}
        \bigoplus_{\ell=1}^6u_j^{(\ell)}=0.
        \label{eq:globalcancel}
    \end{equation}
\end{lemma}
\begin{proof}
    We also need the cancellation for outputs that see no question bit. The
    support column of Table~\ref{tab:hard} shows that the six rows pair identically on
    each interval between consecutive anchors:
    \begin{center}
        \small
        \begin{tabular}{@{}c c@{}}
        \toprule
        Interval & Identical pairs of kernel vectors on that interval\\
        \midrule
        $[0,L]$ & $(1,2),\ (3,4),\ (5,6)$\\
        $[L,2L]$ & $(1,2),\ (3,4),\ (5,6)$\\
        $[2L,3L]$ & $(1,4),\ (2,3),\ (5,6)$\\
        $[3L,4L]$ & $(1,4),\ (2,3),\ (5,6)$\\
        \bottomrule
        \end{tabular}
    \end{center}
    Every pair cancels pointwise and yields the desired equality. 
\end{proof}

\begin{lemma}
    \label{lem:localcancel}
    If $j\in N_a$, we have 
    \begin{equation}
        \bigoplus_{\ell:q_a^{(\ell)}=b}u^{(\ell)}_j=0,\qquad a=0,\ldots,4,\quad b\in\bits,
        \label{eq:localcancel}
    \end{equation}
\end{lemma}
\begin{proof}
    Explicitly, using the results listed in Table~\ref{tab:hard}, one can check:
    \begin{itemize}
        \item For $a=0$, when $q_0=0$, the two questions are $00000$ and $00101$. Both kernel vectors are alternating throughout $N_0$, with the same parity, so they cancel pointwise. When $q_0=1$, the four questions are $10001,10100,11011,11111$; all four kernel vectors are identically one throughout $N_0$, so they cancel pointwise.
        
        \item For $a=1$, when $q_1=0$, the questions are $00000,00101,10001,10100$. In $N_1$, the first two kernel vectors have the same alternating pattern, while the last two are identically one. Thus they cancel pairwise. When $q_1=1$, the questions are $11011$ and $11111$; both have the same pattern in $N_1$, namely all ones to the left of $p_1$ and alternating to the right, so they cancel pointwise.
        
        \item For $a=2$, when $q_2=0$, the questions are $00000,10001,11011$. In $N_2$, the vector for $10001$ is identically one, while those for $00000$ and $11011$ are the two opposite alternating patterns. Hence their pointwise XOR vanishes. When $q_2=1$, the questions are $00101,10100,11111$. To the left of $p_2$, the three patterns are respectively alternating, all ones, and the opposite alternating pattern; to the right of $p_2$, they are respectively all ones, alternating, and the opposite alternating pattern. Hence they cancel pointwise on both sides, and therefore throughout $N_2$.
        
        \item For $a=3$, when $q_3=0$, the questions are $00000,00101,10001,10100$. In $N_3$, the vectors for $00000$ and $10100$ have the same alternating pattern, while those for $00101$ and $10001$ are identically one, so they cancel pairwise. When $q_3=1$, the questions are $11011$ and $11111$; both have the same pattern in $N_3$, namely alternating to the left of $p_3$ and all ones to the right, and therefore cancel pointwise.
        
        \item For $a=4$, when $q_4=0$, the questions are $00000$ and $10100$. Both kernel vectors have the same alternating pattern throughout $N_4$, so they cancel pointwise. When $q_4=1$, the questions are $00101,10001,11011,11111$; all four kernel vectors are identically one throughout $N_4$, and hence cancel pointwise.
    \end{itemize}
    This completes the proof.
\end{proof}

Lemma \ref{lem:globalcancel} and Lemma \ref{lem:localcancel} inspire us to consider a class of functions, called \textit{locally-controlled function}. We will encounter this form of functions in Section \ref{sec:proof_classical_hardness}.

\begin{definition}[Locally-controlled function]
    \label{def:LCF}
    Suppose a deterministic function $f: q \mapsto y$, we say it is a \textit{locally-controlled function} it its output bits $y_j = f_j(q)$ satisfies:
    \begin{itemize}
        \item If $j\in N_a$, then $y_j$ can depend on $q_a$, but it is independent of
        every other question bit $q_c$ with $c\ne a$.
        \item If $j\notin N_0\cup\cdots\cup N_4$, then $y_j$ is independent of the
        entire question $q$.
    \end{itemize}
    That is, an output outside all five neighborhoods has one fixed response value, whereas an output in $N_a$ has at most two response values, one for $q_a=0$ and one for $q_a=1$.
\end{definition}

Using Lemma \ref{lem:globalcancel} and Lemma \ref{lem:localcancel}, we immediately obtain:
\begin{theorem}[Deterministic contradiction]
    \label{thm:deterministic contradiction}
    For any locally-controlled function $f$, we let $y = f(q^{(\ell)})$. As long as $4d < L$, the XOR of these six parity equations satisfies
    \begin{equation}
        \label{eq:}
        \bigoplus_{\ell=1}^6 \bigoplus_{j=0}^{n-1} u_j^{(\ell)T}y_j^{(\ell)} \neq \bigoplus_{\ell=1}^6 \lambda^{(\ell)}.
    \end{equation}
    Moreover, the left-hand side equals $0$ and the right-hand side equals $1$.
\end{theorem}
\begin{proof}
    We check the contribution of each output site $j$.

    \begin{itemize}
    
    \item If $j\in N_a$, write $c_{j,a}(b)$ for the output when the local question is $q_a=b$.  Since no other question bit can affect this output, then by Eq.\,\eqref{eq:localcancel} its contribution is
    \begin{equation}
        \bigoplus_{\ell=1}^6u_j^{(\ell)}y_j^{(\ell)} =\bigoplus_{b\in\bits} c_{j,a}(b) \left(\bigoplus_{\ell:q_a^{(\ell)}=b}u_j^{(\ell)}\right) = 0
    \end{equation}

    \item If $j\notin\bigcup_aN_a$, then $y_j^{(\ell)}=c_j$ is the same bit in every row.  Its total contribution is therefore by Eq.\,\eqref{eq:globalcancel}:
    \begin{equation}
        \bigoplus_{\ell=1}^6u_j^{(\ell)}y_j^{(\ell)} =c_j\left(\bigoplus_{\ell=1}^6u_j^{(\ell)}\right)=0
    \end{equation}
    \end{itemize}
    Thus every site contributes zero, so the XOR of the six complete left-hand sides is $0$. 
    However, the right-hand sides have XOR value $1$, as computed in Table~\ref{tab:hard} of Section~\ref{sec:six_hard}:
    \begin{equation}
        \bigoplus_{\ell=1}^6 \lambda^{(\ell)} = 0+1+1+1+0+0=1\pmod 2.
    \end{equation}
   This completes the proof.
\end{proof}

\subsection{Proof of classical hardness}
\label{sec:proof_classical_hardness}

We let
\begin{equation}
    D:=F(\Pi),\qquad \varepsilon:=\TV(D,P_n).
\end{equation}
If $\varepsilon\ge 1/32$, the conclusion is immediate. Hence assume
\begin{equation}
    \varepsilon<\frac{1}{32}.
\end{equation}

Before going to the final proof, we need to point out that every $q$ has nonnegligible probability.
Let $\mu(q):=D(Q=q)$ be the question marginal of $D$. 
According to Eq.\,(\ref{eq:tv-mon-x}),
\begin{equation}
    \left|\mu(q)-\frac{1}{32}\right| = \left| D(Q=q) - P_n(Q=q) \right| \le \TV(D, P_n) = \varepsilon.
\end{equation}
Therefore, for every $q\in\bits^5$,
\begin{equation}
    \mu(q)\ge\frac{1}{32}-\varepsilon>0. \label{eq:fin-0}
\end{equation}

Now, we can eventually prove Theorem \ref{thm:local deterministic}.
\begin{proof}[Proof of Theorem \ref{thm:local deterministic}]
    For $a=0,\ldots,4$ and $b\in\bits$, define the conditional block law
    \begin{equation}
        \nu_{a,b}:=\Law\!\left(R|_{A_a}\ \middle|\ Q_a=b\right), \quad \nu_{\mathrm{out}}:=\Law\!\left(R|_{A_{\mathrm{out}}}\right).
    \end{equation}
    All these laws are well-defined due to Definition \ref{def:AN_neighborhoods}.
    
    Now introduce \textit{coupling seed}, which consists of the following $11$ auxiliary random blocks 
    \begin{equation}
        \omega
        :=\bigl(R_{\mathrm{out}}, R_{0,0},R_{0,1}, R_{1,0},R_{1,1}, R_{2,0},R_{2,1}, R_{3,0},R_{3,1}, R_{4,0},R_{4,1} \bigr).
    \end{equation}
    where we \emph{choose these blocks mutually independently}: $R_{a,b}\sim\nu_{a,b}$ where $a=0,\ldots,4$ and $b\in\bits$; together with $R_{\mathrm{out}}\sim\nu_{\mathrm{out}}$.
    For every $q\in\bits^5$, define a full source configuration $r(\omega,q)$ by placing the common block $R_{\mathrm{out}}$ on $A_{\mathrm{out}}$ and, for each $a$, placing the selected block $R_{a,q_a}$ on $A_a$:
    \begin{equation}
        r(\omega, q) := (R_{\mathrm{out}}, R_{0,q_0}, R_{1,q_1}, R_{2,q_2}, R_{3,q_3}, R_{4,q_4}).
    \end{equation}
    
    By backward locality, $Q_a$ is a deterministic function only of $R_a$.
    Hence the events $\{Q_a=q_a\}$ depend on disjoint independent blocks $\{A_a\}$, and
    conditioning on $Q=q$ preserves the product form:
    \begin{equation}
        \Law(R\mid Q=q)
        =
        \Law(R_{\mathrm{out}})
        \otimes
        \bigotimes_{a=0}^{4}
        \Law(R_a\mid Q_a=q_a).
        \label{eq:conditional-factorization}
    \end{equation}
    This can be shown directly by calculating (assuming $Q_a = f_a(R_a)$ deterministically)
    \begin{equation}
        \mathrm{Pr}(R|Q) = \frac{\mathrm{Pr}(R_{\mathrm{out}},\{R_a\},\{Q_a\})}{\mathrm{Pr}(\{Q_a\})} = \frac{\mathrm{Pr}(R_{\mathrm{out}}) \mathrm{Pr}(\{R_a\}) \mathbf{1}(\{Q_a=f_a(R_a)\})}{\mathrm{Pr}(\{Q_a\})} =\mathrm{Pr}(R_{\mathrm{out}}) \prod_a \mathrm{Pr}(R_a|Q_a).
    \end{equation}
    
    Let $F_Y$ denote the $Y$-component of $F$. By definition the output $y$ is a deterministic function of $(\omega,q)$
    \begin{equation}
        y(\omega,q):=F_Y\bigl(r(\omega,q)\bigr).
    \end{equation}
    We now fix an arbitrary realization of $\omega$ and study the deterministic map $f^{\omega}(q) =  y(\omega,q)$ which labeled by $\omega$. We recognize that $f^{\omega}$ is a locally-controlled function. In fact: 
    \begin{itemize}
        \item If $j\notin\bigcup_a N_a$, then $\Ball{d}{j}$ is disjoint from every $A_a$. Hence all coordinates visible to $Y_j$ lie in $A_{\mathrm{out}}$, and for fixed $\omega$ there is a bit $c_j^\omega$ such that
        \begin{equation}
            f^\omega(q)=c_j^\omega \qquad\text{for every }q\in\bits^5. \label{eq:c_omega_j}
        \end{equation}
        \item If $j\in N_a$, then $\Ball{d}{j}$ cannot intersect any $A_c$ with $c\ne a$. Indeed, if $\Ball{d}{j}\cap A_c\ne\varnothing$, then $|j-p_c|\le2d$, so $j\in N_c$, contradicting the pairwise disjointness of the $N_a$. Therefore $Y_j$ can see only the common outside block $R_{\mathrm{out}}$ and the single selected block $R_{a,q_a}$. For fixed $\omega$ there is thus a deterministic function $c_{j,a}^\omega:\bits\to\bits$ such that
        \begin{equation}
            f^\omega(q)=c_{j,a}^\omega(q_a) \qquad\text{for every }q\in\bits^5. \label{eq:c_omega_j_a}
        \end{equation}
    \end{itemize}
    
    We now restrict Eq.\,(\ref{eq:c_omega_j})--Eq.\,(\ref{eq:c_omega_j_a}) to the six hard questions $\{q^{(\ell)}\}_{\ell \in [6]}$. 
    For fixed $\omega$, we write $y^{(\ell)}:=f^\omega(q^{(\ell)})$.
    According to Theorem \ref{thm:deterministic contradiction}, we have
    \begin{align}
        \bigoplus_{\ell=1}^6u_j^{(\ell)}y_j^{(\ell)} = c_j^\omega\left(\bigoplus_{\ell=1}^6u_j^{(\ell)}\right)=0, \quad & \text{if $j\notin\bigcup_aN_a$,} \\
        \bigoplus_{\ell=1}^6u_j^{(\ell)}y_j^{(\ell)} = \bigoplus_{b\in\bits} c_{j,a}^\omega(b)
        \left(
            \bigoplus_{\ell:\,q_a^{(\ell)}=b}u_j^{(\ell)}
        \right)
        =0, \quad & \text{if $j\in N_a$.}
    \end{align}

    Therefore
    \begin{equation}
        \bigoplus_{\ell=1}^6 \bigoplus_{j=0}^{n-1}u_j^{(\ell)}y_j^{(\ell)} = 0 .
    \end{equation}
    But Theorem \ref{thm:deterministic contradiction} also claims that the required target parities have odd XOR,
    \begin{equation}
        \bigoplus_{\ell=1}^6\lambda^{(\ell)} =0\oplus1\oplus1\oplus1\oplus0\oplus0=1.
    \end{equation}
    Thus, for every fixed coupling seed $\omega$, at least one of the six hard-question outputs violates its target parity equation.
    If we define
    \begin{equation}
        I_\ell(\omega)
        :=\mathbf 1\!\left[
           \bigoplus_j u_j^{(\ell)}y_j(\omega,q^{(\ell)})\ne\lambda^{(\ell)}
        \right],
    \end{equation}
    then the pointwise contradiction says $\sum_{\ell=1}^6I_\ell(\omega)\ge1$ for every $\omega$.
    Averaging $I_\ell(\omega)$ over $\omega$, we have
    \begin{equation}
        \eta_\ell := \mathbb E_\omega I_\ell(\omega) = D\!\left[ \bigoplus_j u_j^{(\ell)}Y_j\ne\lambda^{(\ell)} \ \middle|\ Q=q^{(\ell)} \right].
    \end{equation}
    Hence,
    \begin{equation}
        \sum_{\ell=1}^6\eta_\ell\ge1. \label{eq:fin-1}
    \end{equation}
    Finally, let $\Win := \{(q, y): P_q(y)>0 \}$ be the support relation of $P_n$. 
    According to Eq.\,(\ref{eq:tv-mon-margin}),
    \begin{equation}
        D(\Win^c) = | D(\Win^c) - P_n(\Win^c) | \le \TV(D, P_n) = \varepsilon.\label{eq:fin-2}
    \end{equation}
    For a hard question $q^{(\ell)}$, violating its parity equation means leaving the support of the target conditional distribution. Therefore, using Eq.\,(\ref{eq:fin-2}), Eq.\,(\ref{eq:fin-0}) and Eq.\,(\ref{eq:fin-1}),
    \begin{equation}
        \varepsilon\ge D(\Win^c) \ge \sum_{\ell=1}^6 \mu(q^{(\ell)})\eta_\ell \ge \left(\frac{1}{32}-\varepsilon\right) \sum_{\ell=1}^6\eta_\ell\ge \frac{1}{32}-\varepsilon
    \end{equation}
    It yields
    \begin{equation}
        \TV(D, P_n) = \varepsilon\ge\frac{1}{64}.
    \end{equation}
    This proves the theorem.
\end{proof}

\subsection{Extension to local noisy classical circuits}
\label{sec:noisy}

We now extend the deterministic lower bound to circuits of local stochastic channels, with the private randomness convention introduced in Sec.~\ref{subsec:app-classical-source-locality}.

\begin{theorem}[Noisy classical generative hardness, Theorem \ref{thm:classical-sampling-hardness} in main text]
    \label{thm:noisy-generative}
    Let $R\sim\Pi$ be an arbitrary product classical source, and let $\mathcal C$ be a circuit of local stochastic channels whose input-output lightcones have bidirectional geometric radius at most $d$. If $4(d+1)<L$, then
    \begin{equation}
        \TV\bigl(\mathcal C(\Pi),P_n\bigr)\ge\frac{1}{64}.
        \label{eq:noisy-generative-gap}
    \end{equation}
\end{theorem}

\begin{proof}
    As explained in Sec.~\ref{subsec:app-classical-source-locality}, expose the independent private randomness of every stochastic gate and regard these variables as additional local source variables. Together with $R$, they form an enlarged product source $\widehat\Pi$, and the noisy circuit becomes a deterministic map $F$ with the same input-output lightcones, such that
    \begin{equation}
        \mathcal C(\Pi)=F(\widehat\Pi).
    \end{equation}
    Theorem~\ref{thm:local deterministic} therefore applies directly to $F$, giving Eq.~\eqref{eq:noisy-generative-gap} whenever $4(d+1)<L$.
\end{proof}

\section{Lemmas for general nonlocal-game construction and transport advantage}
\label{app:general-nonlocal-game}

In this section, we prove the properties of the general nonlocal-game construction introduced in Sec.~\ref{sec:nonlocal-game-transport}. Throughout, all binary vector operations and inner products are understood modulo two. We consider a fixed-dimensional lattice $\Lambda_L$ containing $n$ data qubits and a constant number of marked question sites $p_1,\ldots,p_m$. For each question string $q$, let $C_q$ be a unitary acting on the data register. We assume that the full controlled unitary $U=\sum_q|q\rangle\langle q|\otimes C_q$ admits a constant-depth geometrically local implementation with lightcone radius $R=O(1)$. In particular, both the question and data registers are included when defining geometric lightcones.

For each $q$, recall the subspace $K_q=\left\{u\in\{0,1\}^n:C_qZ^uC_q^\dagger\text{ is diagonal}\right\}$.
Let $\chi=\dim K_q$ and choose a basis $u^{(1)},\ldots,u^{(\chi)}$ of $K_q$. We define $D_{q,\ell}:=C_qZ^{u^{(\ell)}}C_q^\dagger$ and the Boolean functions $h_{q,\ell}$ through $D_{q,\ell}|y\rangle=(-1)^{h_{q,\ell}(y)}|y\rangle$.
The corresponding winning set and conditional distributions are $W_q=\left\{y\in\{0,1\}^n:h_{q,\ell}(y)=0,\ \forall\ell\right\}$. And the distributions are defined as
\begin{equation}
    Q_q=\operatorname{Unif}(K_q^\perp),\qquad P_q=\operatorname{Unif}(W_q).
\label{eq:app-game-distributions}
\end{equation}
We first establish that these definitions are well posed.

\begin{lemma}[Well-definedness of the generalized parity constraints]
\label{lem:app-game-well-defined}
For any unitary $C_q$, the set $K_q$ is a binary linear subspace. The generalized parity constraints defining $W_q$ are mutually consistent and independent of the choice of basis of $K_q$. Moreover, $|W_q|=2^{n-\chi}$, so $P_q$ is a well-defined uniform distribution.
\end{lemma}

\begin{proof}
Since $C_qIC_q^\dagger=I$, the zero vector belongs to $K_q$. For any $u,v\in K_q$, both $C_qZ^uC_q^\dagger$ and $C_qZ^vC_q^\dagger$ are diagonal. Using $Z^{u+v}=Z^uZ^v$, their product is $C_qZ^{u+v}C_q^\dagger$, which is also diagonal. Hence $u+v\in K_q$, proving that $K_q$ is a linear subspace.

Each $D_{q,\ell}$ is a diagonal Hermitian unitary satisfying $D_{q,\ell}^2=I$, and all these operators commute. Consider the joint projector
\begin{equation}
\Pi_q:=\prod_{\ell=1}^{\chi}\frac{I+D_{q,\ell}}{2}
=C_q\left(\prod_{\ell=1}^{\chi}\frac{I+Z^{u^{(\ell)}}}{2}\right)C_q^\dagger.
\label{eq:app-game-projector}
\end{equation}
The first expression shows that $\Pi_q$ is the diagonal projector onto the computational-basis states satisfying all constraints $h_{q,\ell}(y)=0$. Since the basis vectors generate $K_q$, expanding the product gives
\begin{equation}
\Pi_q=2^{-\chi}\sum_{u\in K_q}C_qZ^uC_q^\dagger.
\label{eq:app-game-projector-expansion}
\end{equation}
Taking the trace and using $\operatorname{Tr}(Z^u)=0$ for every nonzero $u$, we obtain
\begin{equation}
|W_q|=\operatorname{Tr}\Pi_q
=2^{-\chi}\sum_{u\in K_q}\operatorname{Tr}(Z^u)
=2^{n-\chi}.
\label{eq:app-game-winning-cardinality}
\end{equation}
Thus $W_q$ is nonempty and contains exactly $2^{n-\chi}$ elements. Furthermore, Eq.~(\ref{eq:app-game-projector-expansion}) depends only on $K_q$, not on its chosen basis. Therefore the projector $\Pi_q$, and consequently its computational-basis support $W_q$, are basis independent.
\end{proof}

We next construct an explicit classical sampler for $Q_q$. The key observation is that $K_q^\perp$ can be generated by local binary vectors extracted from the Heisenberg evolution of single-site Pauli operators.

\begin{lemma}[Local generators of $K_q^\perp$]
\label{lem:app-game-generators}
For each data site $j$, define $O_{q,j}:=C_q^\dagger Z_jC_q$ and expand it in the Hermitian Pauli basis $\mathsf P(a,b)=i^{a^Tb}X^aZ^b$ as
\begin{equation}
O_{q,j}=\sum_{a,b}c_{q,j}(a,b)\mathsf P(a,b).
\label{eq:app-game-Pauli-expansion}
\end{equation}
Let $\mathcal A_{q,j}:=\{a\in\{0,1\}^n:\exists b,\ c_{q,j}(a,b)\neq0\}$. Then
\begin{equation}
K_q^\perp=\operatorname{span}\left(\bigcup_j\mathcal A_{q,j}\right).
\label{eq:app-game-generator-span}
\end{equation}
Consequently, a uniform distribution on $K_q^\perp$ can be generated by taking independent unbiased binary coefficients and forming a linear combination of the vectors in $\bigcup_j\mathcal A_{q,j}$.
\end{lemma}

\begin{proof}
An operator is diagonal in the computational basis if and only if it commutes with every $Z_j$. Therefore, using unitary conjugation,
\begin{equation}
u\in K_q
\iff [Z^u,O_{q,j}]=0,\qquad\forall j.
\label{eq:app-game-K-commutation}
\end{equation}
The Hermitian Pauli operators obey $Z^u\mathsf P(a,b)Z^u=(-1)^{u^Ta}\mathsf P(a,b)$. Substituting Eq.~(\ref{eq:app-game-Pauli-expansion}) gives
\begin{equation}
    [Z^u,O_{q,j}]Z^u
=\sum_{a,b}c_{q,j}(a,b)\left[(-1)^{u^Ta}-1\right]\mathsf P(a,b).
\label{eq:app-game-commutator-expansion}
\end{equation}
Since the Pauli operators are linearly independent, the right-hand side vanishes if and only if $u^Ta=0$ for every pair $(a,b)$ with $c_{q,j}(a,b)\neq0$. The condition depends only on $a$, so the index $b$ can be eliminated using the definition of $\mathcal A_{q,j}$. Applying this condition for every $j$, we obtain
\begin{equation}
K_q=\left\{u:u^Ta=0,\ \forall a\in\bigcup_j\mathcal A_{q,j}\right\}
=\left(\operatorname{span}\left(\bigcup_j\mathcal A_{q,j}\right)\right)^\perp.
\label{eq:app-game-K-span}
\end{equation}
Taking the orthogonal complement of both sides and using $(V^\perp)^\perp=V$ for binary vector spaces proves Eq.~(\ref{eq:app-game-generator-span}).

To establish the sampling statement, arrange the vectors in $\bigcup_j\mathcal A_{q,j}$ as the columns of a binary matrix $G_q$, allowing repeated columns. Equation~(\ref{eq:app-game-generator-span}) implies $\operatorname{im}G_q=K_q^\perp$. If $r$ consists of independent unbiased bits, then $G_qr$ is uniform on $\operatorname{im}G_q$. Indeed, for any $z\in\operatorname{im}G_q$, the solution set of $G_qr=z$ is a coset of $\ker G_q$ and therefore contains exactly $|\ker G_q|$ elements, independently of $z$. Hence $G_qr\sim\operatorname{Unif}(K_q^\perp)=Q_q$.
\end{proof}

The preceding lemma proves the sampling identity but does not yet establish geometric locality. For this purpose, we use a fixed collection of local random variables, whose spatial positions and number do not depend on the question string.

\begin{lemma}[Geometrically local classical sampling]
\label{lem:app-game-local-sampler}
Suppose the full controlled unitary $U$ has geometric lightcone radius at most $R$. Then the source distribution $Q(q,z)=\pi(q)Q_q(z)$ can be sampled exactly from independent local randomness by a classical circuit of bidirectional geometric radius at most $2R$, with depth independent of the system size.
\end{lemma}

\begin{proof}
For each site $j$, define $\mathcal I_j:=\{a\in\{0,1\}^n:\operatorname{supp}(a)\subseteq B_R(j)\}$. We first show that $\mathcal A_{q,j}\subseteq\mathcal I_j$ and that membership in $\mathcal A_{q,j}$ depends only on question variables within $B_R(j)$.

Consider the operator on the full question and data registers,
\begin{equation}
\widetilde O_j:=U^\dagger(I\otimes Z_j)U
=\sum_q|q\rangle\langle q|\otimes O_{q,j}.
\label{eq:app-game-controlled-Heisenberg}
\end{equation}
By the geometric locality of $U$, the operator $\widetilde O_j$ is supported within $B_R(j)$. Consequently, each conditional operator $O_{q,j}$ is supported on the data qubits in this neighborhood. Its Pauli expansion can therefore contain only terms with $\operatorname{supp}(a),\operatorname{supp}(b)\subseteq B_R(j)$, proving $\mathcal A_{q,j}\subseteq\mathcal I_j$.

Furthermore, because $\widetilde O_j$ acts trivially on question qubits outside $B_R(j)$, two question strings $q$ and $q'$ that agree within this neighborhood must satisfy $O_{q,j}=O_{q',j}$. Thus the coefficients $c_{q,j}(a,b)$, and in particular the sets $\mathcal A_{q,j}$, depend only on the local restriction $q|_{B_R(j)}$.

For every pair $(j,a)$ with $a\in\mathcal I_j$, introduce an independent unbiased random bit $r_{j,a}$ located at site $j$. Define
\begin{equation}
g_{q,j,a}:=\mathbf 1[a\in\mathcal A_{q,j}],
\qquad
(G_q)_{i,(j,a)}:=g_{q,j,a}a_i.
\label{eq:app-game-G-definition}
\end{equation}
The column indices $(j,a)$ are fixed independently of $q$, while the binary coefficients $g_{q,j,a}$ depend only on $q|_{B_R(j)}$. The sampler outputs
\begin{equation}
z_i=(G_qr)_i
=\bigoplus_{j, a\in\mathcal I_j}g_{q,j,a}r_{j,a}a_i.
\label{eq:app-game-sampling-rule}
\end{equation}
By Lemma~\ref{lem:app-game-generators}, the nonzero columns of $G_q$ span exactly $K_q^\perp$. Hence $\operatorname{im}G_q=K_q^\perp$, and Eq.~(\ref{eq:app-game-sampling-rule}) samples $Q_q$ exactly.

It remains to prove locality. If $(G_q)_{i,(j,a)}\neq0$, then $a_i=1$, and therefore $i\in B_R(j)$. Thus the random variable $r_{j,a}$ can affect only outputs within distance $R$ of its location $j$. Conversely, the output $z_i$ depends only on random variables located at sites $j\in B_R(i)$.

The dependence on the question string is also local. Since $g_{q,j,a}$ depends only on $q|_{B_R(j)}$, a question bit at position $p_\alpha$ can affect $z_i$ only if there exists a site $j$ satisfying $\operatorname{dist}(j,p_\alpha)\le R$ and $\operatorname{dist}(i,j)\le R$. The triangle inequality then gives $\operatorname{dist}(i,p_\alpha)\le2R$. Therefore every output has a backward lightcone of radius at most $2R$, and every question bit has a forward lightcone of radius at most $2R$. The forward lightcones of the independent random bits have radius at most $R$. This establishes bidirectional geometric locality.

Finally, because the lattice has fixed dimension and bounded local degree, $s_R:=\max_j|B_R(j)|=O(1)$. Each site requires at most $|\mathcal I_j|=2^{|B_R(j)|}\le2^{s_R}$ independent random bits. Moreover, each output combines contributions from at most $s_R2^{s_R}=O(1)$ local random variables. The coefficients $g_{q,j,a}$ can be specified by finite lookup tables indexed only by nearby question variables. Hence the sampler can be implemented by a constant-depth classical circuit with bounded-range gates. The question bits themselves are sampled independently from $\pi$ at their marked sites and are left unchanged.
\end{proof}

We now establish exact quantum transport. 

\begin{lemma}[Exact unitary transport]
\label{lem:app-game-transport}
For any unitary family $C_q$, the distributions defined in Eq.~(\ref{eq:app-game-distributions}) satisfy
\begin{equation}
    C_q\operatorname{diag}(Q_q)C_q^\dagger
    =\operatorname{diag}(P_q).
    \label{eq:app-game-exact-conditional}
\end{equation}
\end{lemma}

\begin{proof}
Since $Q_q$ is uniform on $K_q^\perp$, the character orthogonality relation over binary vector spaces gives
\begin{equation}
    \operatorname{diag}(Q_q)=2^{-n}\sum_{u\in K_q}Z^u.
    \label{eq:app-game-Q-Fourier}
\end{equation}
Meanwhile, Lemma~\ref{lem:app-game-well-defined} shows that $P_q$ is uniform on $W_q$, which contains $2^{n-\chi}$ elements. The projector $\Pi_q$ in Eq.~(\ref{eq:app-game-projector}) has computational-basis support exactly $W_q$. Therefore,
\begin{equation}
    \operatorname{diag}(P_q) = 2^{\chi-n}\Pi_q = 2^{\chi-n}\prod_{\ell=1}^{\chi}\frac{I+D_{q,\ell}}{2} = 2^{-n}\sum_{u\in K_q}C_qZ^uC_q^\dagger.
    \label{eq:app-game-P-Fourier}
\end{equation}
Comparing Eqs.~(\ref{eq:app-game-Q-Fourier}) and (\ref{eq:app-game-P-Fourier}) immediately gives Eq.~(\ref{eq:app-game-exact-conditional}).
\end{proof}